\documentclass[aps,prb,twocolumn,floatfix,showpacs,showkeys]{revtex4}
\usepackage{tabularx}
\usepackage{bm}
\usepackage{euscript}
\usepackage{epsfig,psfrag,subfigure,subeqnarray}
\usepackage{graphicx}
\usepackage{color}
\usepackage{amsfonts}
\usepackage{exscale}
\usepackage{amsmath}
\usepackage{amsbsy}
\usepackage{dcolumn}

\def\lan{\langle}
\def\ran{\rangle}
\def\llan{\langle\langle}
\def\rran{\rangle\rangle}

\def\va{\varepsilon}
\def\sa{\sigma}

\def\bsa{{\bar{\sigma}}}
\def\kp{{ k^\prime}}
\def\ka{{ k\sa}}
\def\kb{{ \kp\sa}}
\def\kpp{{ k^{\prime\prime }}}

\def\wa {\omega}
\newcommand{\bd}{\begin{equation}}
\newcommand{\ed}{\end{equation}}
\newcommand{\be}{\begin{equation}}
\newcommand{\ee}{\end{equation}}
\newcommand{\bt}{\begin{split}}
\newcommand{\et}{\end{split}}

\newcommand{\bn}{\begin{align}}
\newcommand{\en}{\end{align}}

\newcommand{\bea}{\begin{eqnarray}}
\newcommand{\eea}{\end{eqnarray}}
\newcommand{\ba}{\begin{array}}
\newcommand{\ea}{\end{array}}
\newcommand{\nn}{\nonumber}

\newcommand{\bra}{\langle\langle }
\newcommand{\ket}{\rangle\rangle }

\begin{document}

\title{Anderson Model out of equilibrium: decoherence effects in transport through a quantum dot}
\author{Rapha\"{e}l Van Roermund, Shiue-yuan Shiau and Mireille Lavagna$^*$}
\affiliation{Commissariat \`{a} l'Energie Atomique de Grenoble INAC/SPSMS, 17 rue des Martyrs, 38054 Grenoble Cedex 9, France}
\date{\today}

\begin{abstract}
The paper deals with the nonequilibrium two-lead Anderson model, considered as an adequate description for transport through a d-c biased quantum dot. Using a self-consistent equation-of-motion method generalized out of equilibrium, we calculate a fourth-order decoherence rate
$\gamma^{(4)}$ induced by a bias voltage $V$. This decoherence rate provides a cut-off to the infrared divergences of the self-energy showing up in the Kondo regime. At low temperature, the Kondo peak in the density of states is split into two peaks pinned at the chemical potential of the two leads. The height of these peaks is controlled by $\gamma^{(4)}$. The voltage dependence of the differential conductance exhibits a zero-bias
peak followed by a broad Coulomb peak at large $V$, reflecting charge fluctuations inside the dot. The low-bias differential conductance is found to be a universal function of the normalized bias voltage $V/T_K$, where $T_K$ is the Kondo temperature. The universal scaling with a single energy scale $T_K$ at low bias voltages is also observed for the renormalized decoherence rate $\gamma^{(4)}/T_K$. We discuss the effect of $\gamma^{(4)}$ on the crossover from strong to weak coupling regime when either the temperature or the bias voltage is increased.
\end{abstract}

\maketitle

\section{Introduction}

Over the last ten years, an intense experimental and theoretical
activity has been developed to study quantum dots. Due to the
presence of strong electronic correlations in the dot, these
mesoscopic systems give rise to rich collective phenomena such as
the Coulomb blockade and the Kondo effect. Their manifestations in transport can be studied in a detailed and controlled way in
such devices as semiconductor-based quantum dots embedded in a
two-dimensional electron gas\cite{Goldhaber1998} or carbon
nanotubes\cite{Nygard00}. One of the great interests of these
systems is to offer the possibility of studying them under
nonequilibrium conditions when either a bias voltage is applied to
the leads or an electromagnetic field irradiates the device.\

A simple model describing quantum dots is the Anderson
model\cite{Anderson61}, in which the dot is represented by a
localized level connected to Fermi seas of conduction electrons
through tunneling barriers. When the dot is singly occupied, it
has been shown that the linear conductance increases as one lowers
the temperature, and eventually reaches the unitary limit $2e^2/h$
in the case of symmetric coupling to the leads. This was predicted
in the context of quantum dots twenty years
ago\cite{Glazman88,Ng88} and was observed experimentally about ten
years later\cite{Goldhaber1998,Nygard00}.

While in equilibrium most of the properties of the Kondo effect
are now well understood\cite{Hewson} thanks to the development of a panel of powerful techniques (e.g.
renormalization group, 
Bethe ansatz, 
Fermi-liquid theory, 
conformal field theory, 
density matrix renormalization group, 
slave boson and equation-of-motion approaches),
most of these techniques fail out of equilibrium. Hence there is a
huge interest to develop new techniques to tackle the problem of
the Kondo effect out of equilibrium and more generally nonequilibrium effects in strongly-correlated electron systems.

Theoretically, the Kondo effect out of equilibrium has been
investigated by a variety of techniques developed most of the time
within the Keldysh formalism: perturbation theory and perturbative
renormalization group approach\cite{Kaminski00,Coleman2001,Parcollet02,Rosch2001,Rosch2003,Rosch2004}, slave-boson formulation solved by
using either mean-field\cite{Aguado00} or non-crossing
approximation\cite{WingreenPRB94}, equation-of-motion
approaches\cite{Monreal2005,Swirkowicz03}. Exact solutions at the Toulouse limit have been proposed\cite{Schiller95}. Other ones
have extended the Bethe ansatz out of
equilibrium\cite{Konik01,Konik02}, and in some cases have used the
results to construct a Landauer-type picture of transport through the
quantum dot. There have also been important efforts to develop numerical techniques such as time-dependent Numerical Renormalisation Group\cite{Bulla08,Anders2008} and imaginary-time theory solved by using Quantum Monte Carlo\cite{HanHeary2007}. All
those approaches have only a limited validity of their parameter
regimes since they mostly describe the properties of the system in
its ground state, and not in its excited many-body states reached
when the bias voltage drives a current through the dot.

In this paper, we develop an equation-of-motion (EOM) approach to
tackle the nonequilibrium Kondo effect in quantum dots and study
the decoherence effects induced by a bias voltage. The EOM method, though conceptually simple, requires some care. The
recursive application of the Heisenberg equation of
motion\cite{Zubarev60} generates an infinite hierarchy of
equations, which relate the different Green functions of the
system. This hierarchy has to be truncated by a suitable
approximation scheme in order to form a closed set of equations. The choice of the truncation
scheme is crucial in order to treat carefully the correlation
effects both from the Coulomb interaction and from the dot-lead
tunneling.

The EOM technique was applied to the original Anderson model
at equilibrium a long time ago\cite{Dworin1967,Appelbaum1969,Theumann69etal,Lacroix1981} in the context of the dilute
magnetic alloys. When applying the standard approximation
based on a truncation of the equations of motion at second order
in the hybridization term $t_\sigma$, it yields results which
agree with perturbation theory calculations for temperatures above
the Kondo temperature, $T_K$. This truncation scheme is usually referred to as the Lacroix approximation\cite{Lacroix1981}. Even though the scheme has serious
drawbacks at this level of approximation (underestimation of the
Kondo temperature $T_K$, absence of Kondo effect just at the
particle-hole symmetric point), it is acknowledged to provide a
valuable basis for the description of the Kondo effect both at
high and low temperatures. The applicability of the Lacroix approximation is
nicely reported in a recent paper by V. Kashcheyevs {\it et
al}\cite{Kashcheyevs2006}.

In the early nineties, Meir, Wingreen, and Lee\cite{MWL1993} undertook to apply
the EOM method to the study of quantum dots out of equilibrium and/or in the presence of a magnetic
field. They used a simplified version of the Lacroix approximation, which fails to
account for the finite decoherence rates induced by bias voltage
and/or magnetic field. Meir {\it et al.} proposed to
introduce them heuristically by making use of
the Fermi golden rule. They obtained interesting results for the
bias voltage dependence of the differential conductance showing a zero-bias anomaly, but the weakness of the approach is that it
does not constitute an unified and consistent frame for the
treatment of the Kondo effect in the presence of the decoherence
effects induced out of equilibrium.

There have been recent attempts to use an
approximation which truncates the equations of motion at higher
order in $t_\sigma$ \cite{Monreal2005,Luo1999,Swirkowicz03}. Their authors claimed to improve
quantitatively at equilibrium the Kondo temperature and the
density of states around the Fermi energy and have been able to investigate some nonequilibrium issues. However, there is
need to clarify the decoherence effects in the framework of the EOM method.

The organization of the paper is the following:

- In Sec.\ref{secEOM}, we outline the EOM formalism and the main
steps of the proposed approximation based on a truncation
of the equations of motion at the fourth order in $t_\sigma$. An
analytical expression of the retarded Green function in the dot is
derived involving expectation values which are determined
self-consistently. The details of the calculations are presented
in \ref{App:EOM} and \ref{App:TA1}.

- An analytical study of this Green function is presented in
Sec.\ref{sec:analytic}. Namely, we deduce the renormalization
effects and the transition rates involved at the second order in
$t_\sigma$ for the different regimes of the Anderson model. Special care is given to the
singly-occupied dot regime, for which the vanishing of one of the
transition rates leads to the low-energy logarithmic divergence of
the self-energy of the dot Green function, yielding a Kondo resonance peak in the density of states. We
show how the approximation scheme allows one to derive a decoherence rate out of equilibrium which becomes finite as soon as a
bias voltage is applied. This decoherence rate provides a cut-off
to the logarithmic divergence present at equilibrium in the Kondo
regime. We show how our approximation scheme improves the result for the
Kondo temperature $T_K$ upon earlier predictions.

- We present in Sec.\ref{secResult} our numerical results - both at
equilibrium and out of equilibrium - for the density of states, the
(linear and differential) conductance and the bias-induced
decoherence rate. A self-consistent treatment is required in order to determine
the expectation values involved in the Green function of the dot.
At equilibrium, the density of states in the particle-hole
symmetric case shows a three-peak structure at low temperature
with a Kondo resonance peak in the local moment regime. This
result constitutes an advantage of our approximation scheme
compared to the Lacroix approximation. The unitary limit for the linear
conductance $G=2e^2/h$ is analytically recovered at zero
temperature in the particle-hole symmetric case when the dot is
symmetrically coupled to the leads. Numerically, one notices a
slight underestimation of $G$ due to the numerical accuracy of the self-consistent treatment.

Out of equilibrium, the spectral function shows a splitting of the Kondo peak into two peaks pinned at the chemical potentials of the two leads.  The height of the peaks diminishes when the bias voltage increases, meaning that the Kondo effect is destroyed by decoherence induced out of equilibrium. This influences the evolution of the differential conductance as a function of bias voltage, which shows a zero-bias anomaly followed by a broad Coulomb
peak. At low bias voltage, we check that the differential conductance follows a universal scaling law
which depends on a single energy scale, $T_K$. Comparison is made
with results obtained by recent numerical techniques for
nonequilibrium such as time-dependent NRG\cite{Anders2008} and
imaginary time theory solved by QMC\cite{HanHeary2007,Han2009}.

Finally, we compute the bias voltage dependence of the decoherence
rate and discuss the crossover from strong coupling to weak coupling
regime depending on the comparison between the decoherence rate and a characteristic energy
scale $T^*$. We show the existence of a
crossover between the strong coupling and weak coupling regime
when either the temperature or the bias voltage is raised, as
predicted by other methods.

\section{Equation-of-motion formalism\label{secEOM}}

We model the quantum dot connected to the two leads by the single-level (spin-1/2) Anderson impurity hamiltonian
\bea
  \mathcal{H}&=&H_{lead}+H_{dot}+H_{int}, \label{AndersonModel}\\
  H_{lead}&=&\displaystyle \sum_{\alpha k\sa } \varepsilon_{\alpha k} c_{\alpha k\sa}^{\dag}c_{\alpha k\sa},\nn\\
 H_{dot}&=& \sum_{\sa}\va_{\sa}n_{\sa}+U n_{\uparrow}n_{\downarrow},\nn \\
 H_{int} &=&\displaystyle \sum_{\alpha k\sa } (t_{ \alpha \sa}c^{\dag}_{\alpha k\sa}f_{\sa} + H.C.)\nn ,
 \eea
where $c_{\alpha k\sa }^{\dag }(c_{\alpha k\sa })$ is the creation (annihilation) operator of an electron of momentum $k$ and spin $\sigma(=\pm 1)$ in the $\alpha(= L,~R)$ lead (with energy $\varepsilon_{\alpha k}=\va_k-\mu_\alpha)$; $\mu_\alpha$ is the chemical potential in the $\alpha$ lead; $f_{\sigma }^{\dag}(f_{\sigma })$ is the creation (annihilation) operator of an electron of spin $\sigma$ in the quantum dot (with energy $\varepsilon _{\sigma}=\va_d+\sigma B/2$ when the Zeeman splitting $B$ is included); $n_\sa=f_\sa^\dag f_\sa$ is the number operator for electrons of spin $\sigma$ in the dot; $U$ is the Coulomb interaction between two electrons of opposite spin in the dot; and $t_{\alpha \sa}$ is the tunneling matrix element between the state $|k \sa\rangle$ in the $\alpha$ lead, and the state $|\sa\rangle$ in the dot. For simplicity, we assume $t_{\alpha\sa}$ to be real and $k$-independent.\

When a bias voltage is applied to the leads
$(eV=\mu_L-\mu_R)$, the system is driven out of equilibrium
and a current is induced through the quantum dot. The current $I$ for the Anderson model is expressed by the generalized Landauer formula\cite{MW92} accounting for the interactions among electrons 
\bd I=\frac{2e}{\hbar}\sum_\sa
\int_{-W}^W d\va\frac{\Gamma_{L\sa }(\va)\Gamma_{R\sa
}(\va)}{\Gamma_{L \sa }(\va)+\Gamma_{R\sa
}(\va)}[f_F^L(\va)-f_F^R(\va)]\rho_\sa(\va) , \label{Landauer} 
\ed
where 
\begin{itemize}
	\item $W$ is the half-bandwidth of the conduction electron band in the leads,
	\item $\Gamma_{\alpha \sa}(\va)$ is the tunneling rate of the spin $\sa$  dot electron at energy $\va$ into the lead $\alpha$, defined as $\Gamma_{\alpha\sa}(\va)=\pi \sum_{k} t_{\alpha\sa}^2 \delta(\va-\va_{\alpha k})=\pi t_{\alpha\sa}^2 \rho_{\alpha}^{0}(\va)$ with $\rho_{\alpha}^{0}(\va)$ the unrenormalized density of states at energy $\va$ in the lead $\alpha$,
	\item $f_F^\alpha(\va)= \{{\rm exp}\left[\beta(\va-\mu_\alpha)\right]+1\}^{-1}$ is the Fermi-Dirac distribution function in the $\alpha$ lead,
	\item $\rho_\sa(\va)$, the local density of states for spin $\sa$ in the dot, can be expressed in terms of the retarded electron Green function in the dot $\mathcal{G}_{\sa}^{r}(\va)$ according to $\rho_\sa(\va)=-1/\pi{\rm Im}\mathcal{G}_{\sa}^{r}(\va)$.
\end{itemize}
As pointed out in Ref.\onlinecite{MW92}, Eq.(\ref{Landauer}) is valid provided that the tunneling couplings for both leads $\Gamma_{L\sa }(\va)$ and $\Gamma_{R\sa}(\va)$ differ only by a constant multiplicative factor.\

The task is to compute the
retarded electron Green function in the dot defined as
$\mathcal{G}_{\sigma}^{r}(\wa) =-i\int_{0}^{\infty } dt e^{i\zeta
t}\langle \{f_{\sigma }(t),f_{\sigma }^{\dag }(0)\}\rangle$
where $\zeta =\wa+i\delta$ ($\delta \rightarrow 0^+$). In order to simplify the notations in the rest of the paper, the imaginary part $i\delta$ going alongside $\wa$ will be implicit, while the summation over $k$ implies
summation over both $\alpha$ and $k$. Hence we write in a
shorthand notation
\[
\sum_{\alpha =L,R}\sum_k t_{\alpha \sa}\longrightarrow \sum_k t_\sa.
\]
Using the Zubarev notation\cite{Zubarev60} for the retarded Green functions involving fermionic operators $A$ and $B$
\begin{equation}
\bra A , B \ket = - i \; \lim_{\delta \rightarrow 0^+} \; \int^\infty_{0} dt \; e^{i(\omega + i\delta)t} \; \langle\{A(t),B(0)\}\rangle  , \label{Zubarov1}
\end{equation}
which can be integrated by parts, and using the Heisenberg equation of motion, one can show the following relation
\bea 
\wa\langle\langle A,B\rangle\rangle = \langle\{A,B\}\rangle +\langle\langle [ A,H ] ,B\rangle\rangle \label{eq:Zubarov} . 
\eea 
This allows us to derive a flow of equations for the dot Green function $\mathcal{G}^r_{\sigma }(\wa) \equiv \langle \langle f_{\sigma },f_{\sigma }^{\dag }\rangle \rangle$.\ 
We also adopt a simpler notation in the following derivations by changing
\[
\llan A,f_{\sa}^{\dag} \rran\longrightarrow \llan A \rran.
\]\

Applying Eq.~(\ref{eq:Zubarov}), we find the first equations of motion
\begin{eqnarray}
(\wa-\varepsilon_{\sa})\langle\langle f_{\sa}\rangle\rangle &=& 1+\sum_{k} t_{\sa}\langle\langle
c_{k\sa}\rangle\rangle +U\langle\langle n_\bsa f_{\sa}\rangle\rangle ,
\label{eq:f1}\\
 (\wa-\va_k )\langle\langle c_{k\sa}\rangle\rangle &=&t_{\sa}\langle\langle
f_{\sa}\rangle\rangle . \label{eq:f2}
\end{eqnarray}
Combining Eqs.~(\ref{eq:f1}, \ref{eq:f2}) yields
\begin{equation}
[\wa-\varepsilon_{\sa}-\Sigma^0_\sa(\wa)]\langle\langle f_{\sa}\rangle\rangle =1 +U\langle\langle n_\bsa f_{\sa}\rangle\rangle ,\label{eq:f3}\\
\end{equation}
where $\Sigma^0_\sa(\wa)=\displaystyle \sum_{k} \frac{t_{\sa}^2}{\wa-\varepsilon_k}$. Eqs.~(\ref{eq:f1},\ref{eq:f2}) are referred to as the first-generation equations of motion in the hierarchy. They govern the evolution of the Green functions formed by a single operator (i.e. $\langle\langle f_{\sa}\rangle\rangle$ and $\langle\langle c_{k\sa}\rangle\rangle$).\

Throughout this paper, we assume that the half-bandwidth $W$ is much
larger than all the other energy scales, so that the band edge effect
does not affect the local density of states in the dot
$\rho_\sa(\wa)$. In this case, the properties of the system at low
temperatures do not depend on the exact value of $W$ since only
states around the Fermi level contribute, justifying the
consideration of the wide-band limit\cite{Appelbaum1969}
$W\rightarrow\infty$. Within this limit, the non-interacting
self-energy can be approximated by $\Sigma^0_\sa(\wa)\simeq
-i\Gamma_\sa$, where $\Gamma_\sa(=\Gamma_{L\sa }+\Gamma_{R\sa })$
is a constant, independent of energy.\

Interesting dynamics comes from $\langle\langle
n_\bsa f_{\sa}\rangle\rangle$; its equation of motion is given by
\begin{eqnarray}
\lefteqn{[\wa-\varepsilon_{\sa} -U]\langle\langle n_\bsa f_{\sa}\rangle\rangle = \lan n_\bsa\ran  +\sum_{k} \Big[t_{\sa}
\langle\langle n_\bsa c_{k\sa}\rangle\rangle}\hspace{2cm}\nn\\
& & +t_\bsa \langle\langle f^{\dag}_\bsa c_{k\bsa}f_{\sa}\rangle\rangle-t_\bsa \langle\langle c_{k\bsa}^{\dag}f_\bsa f_{\sa}\rangle\rangle\Big].  \label{eq:f4}
\end{eqnarray}

The Green functions appearing on the right-hand side of Eq. (\ref{eq:f4}) have their evolution governed by the following equations
\begin{widetext}
\begin{subeqnarray}
\wa_{:k}\langle\langle n_\bsa c_{k\sa}\rangle\rangle &=&t_{\sa}
\langle\langle n_\bsa f_{\sa}\rangle\rangle +\sum_{k^\prime}t_\bsa
\Big[\langle\langle f_\bsa ^{\dag}c_{ k^\prime
\bsa}c_{k\sa}\rangle\rangle -\langle\langle
c_{ k^\prime \bsa}^{\dag}f_\bsa c_{k\sa}\rangle\rangle  \Big],\slabel{eq:f44a} \\
\wa_{\bsa:\sa k} \langle\langle
f_\bsa^{\dag}c_{k\bsa}f_{\sa}\rangle\rangle &=& \langle
f_\bsa^{\dag} c_{k\bsa}\rangle  + t_{\bsa} \langle\langle n_\bsa
f_{\sa}\rangle\rangle +\sum_{ k^\prime}\Big[ t_{\sa}\langle\langle
f_\bsa^{\dag}c_{k\bsa}c_{ k^\prime \sa}\rangle\rangle
-t_{\bsa}\langle\langle
c_{ k^\prime \bsa}^{\dag}c_{k\bsa}f_{\sa}\rangle\rangle \Big],\slabel{eq:f44b}\\
\left(\wa_{k:\sa\bsa} - U\right) \langle\langle
c_{k\bsa}^{\dag}f_\bsa f_{\sa}\rangle\rangle &=& \langle
c_{k\bsa}^{\dag} f_\bsa\rangle  -t_\bsa\langle\langle n_\bsa
f_{\sa}\rangle\rangle +\sum_{ k^\prime} \Big[t_\bsa \langle\langle
c_{k\bsa}^{\dag}c_{ k^\prime \bsa}f_{\sa}\rangle\rangle
+t_{\sa}\langle\langle c_{k\bsa}^{\dag}f_\bsa c_{ k^\prime
\sa}\rangle\rangle \Big]. \slabel{eq:f44c}
\label{eq:f44}
\end{subeqnarray}
\end{widetext}
where we write in a shorthand notation
\bea
\wa_{\alpha \beta \cdots : a b \cdots} &\equiv&
\wa+\va_\alpha+\va_\beta+\cdots-\va_a-\va_b-\cdots  , \label{eq:notation1} \nn
\eea
with $\{\alpha \beta \cdots, a b \cdots\}$ being any set of parameters within ${k}$'s and ${\sa}$'s. We have for instance: $\wa_{:k}\equiv \wa-\va_k $, $\wa_{k:}\equiv \wa +\va_k$, $\wa_{\bsa:k\sa}\equiv \wa +\varepsilon_\bsa-\va_k-\varepsilon_{\sa}$, and $\wa_{k:\sa\bsa} \equiv \wa +\va_k-\varepsilon_{\sa}-\varepsilon_\bsa$. \

Eqs.~(\ref{eq:f44}) generate three new Green functions on their right-hand side. Generally one can divide the expressions of the latter Green functions into two parts\begin{subeqnarray}
\label{eq:f42}
\bra f_\bsa^\dag c_{k\bsa} c_{k'\sa} \ket &=& \lan f_\bsa^\dag c_{k\bsa} \ran    \bra c_{k'\sa} \ket + \bra f_\bsa^\dag c_{k\bsa} c_{k'\sa} \ket_c,  \;\; \\
\slabel{eq:f42b}
\bra c_{k\bsa}^\dag f_\bsa c_{k'\sa} \ket &=& \lan c_{k\bsa}^\dag f_\bsa \ran    \bra c_{k'\sa} \ket + \bra c_{k\bsa}^\dag f_\bsa c_{k'\sa} \ket_c, \;\; \\
\slabel{eq:f42c}
\bra c_{k\bsa}^\dag c_{k'\bsa} f_{\sa} \ket &=& \lan c_{k\bsa}^\dag c_{k'\bsa} \ran    \bra f_{\sa} \ket + \bra c_{k\bsa}^\dag c_{k'\bsa} f_{\sa} \ket_c . \;\;
\end{subeqnarray}
where the first part is obtained by decoupling pairs of same-spin
operators and the second part $\bra\cdots \ket_c$ defines
connected Green functions in the spirit of cumulant expansion. It
is often assumed that these connected Green functions are
negligible. This assumption has been broadly applied, usually referred to as the Lacroix
approximation\cite{Lacroix1981} . It turns out that within this approximation, the calculation
of $\langle\langle f_{\sa}\rangle\rangle$ is exact at the second
order in $t_\sa$, while it picks up some of the fourth-order contributions. At zero temperature, this
approximation leads to logarithmic singularities in the density of states of the dot at the chemical potential, even in the presence
of an external magnetic field $B$ or a bias voltage
$V$. These divergences are unphysical since one expects the
logarithmic singularities to be washed out by decoherence effects
introduced by either $B$ or $V$.\

In this work, we propose to go beyond the Lacroix approximation, and
consider higher hierarchy in the equations of motion. The
interest of the approach is to account for the decoherence effects
introduced by either nonequilibrium or the presence of a magnetic
field. The detailed derivation of the higher-hierarchy equations of motion and
the decoupling scheme are given in \ref{App:EOM}. In the approximation scheme we propose, after having
expanded the equations of motion to order $t_\sa^4$, we decouple pairs of
same-spin lead electron operators (e.g. $\lan c_{\kp \sa}^\dag
c_{k\sa}\ran$) and pairs of same-spin dot-lead electron operators
(e.g. $\lan c_{k \sa}^\dag f_{\sa}\ran$ and $\lan f_{ \sa}^\dag
c_{k\sa}\ran$). This decoupling has to be done carefully in order
to avoid double counting. The equations of motion of the three functions on the
left-hand-side of Eqs.~(\ref{eq:f4}) are given by
Eqs.~(\ref{app:eq:f403}), which are exact up to order $t_\sa^4$. \

Combining Eqs.~(\ref{eq:f3},~\ref{eq:f4},~\ref{app:eq:f403})
yields a rather complex expression for $\bra f_\sa \ket$ since Eqs.~(\ref{app:eq:f403}) couple to
each other in an integral way. In this paper we will limit
ourselves to a simple expression for the Green function $\bra
f_\sa \ket$ that neglects contributions generated through integral
coupling of the equations of motion. This is motivated by the fact that the
contributions we keep effectively lead to a second order term
after resummation, as shown later on. The integral terms we
neglect, on the contrary, are at least of fourth order in
$t_\sigma$ and we believe they are not relevant for our study of
the decoherence effects at nonzero bias and/or temperature.
Thus, we are able to reduce Eqs.~(\ref{app:eq:f403}) to
\begin{widetext}
\begin{subeqnarray}
\label{eq:f405}
\big[\wa_{:k}-\Sigma_{1\sa}(\wa_{:k})\big]\langle\langle n_\bsa c_{k\sa}\rangle\rangle &=&t_{\sa} \langle\langle n_\bsa f_{\sa}\rangle\rangle +\Sigma_{5\bsa}(\wa_{:k}) \langle\langle n_\sa c_{k\sa}\rangle\rangle,\slabel{eq:f4a05}\\
\big[\wa_{\bsa:k\sa}-\widehat{\Sigma}_{2\sa}(\wa_{:k})\big]\langle\langle f_\bsa^{\dag}c_{k\bsa}f_{\sa}\rangle\rangle &=&
\langle f_\bsa^{\dag} c_{k\bsa}\rangle  + t_{\bsa} \langle\langle n_\bsa f_{\sa}\rangle\rangle - \sum_\kp  \Big[  t_\bsa f_{\kp k}^\bsa  +\lan f_\bsa^\dag c_{k\bsa} \ran  t^2_\sa D_{\bsa:k \kp}\sum_{\kpp} f_{\kpp \kp}^\sa\Big] \llan f_\sa\rran  ,\slabel{eq:f4b05}\\
\big[\wa_{k:\sa\bsa}-U-\widehat{\Sigma}_{3\sa}(\wa_{k:})\big]\langle\langle c_{k\bsa}^{\dag}f_\bsa f_{\sa}\rangle\rangle &=&
\langle c_{k\bsa}^{\dag} f_\bsa\rangle  -t_\bsa\langle\langle
n_\bsa f_{\sa}\rangle\rangle + \sum_\kp \Big[  t_\bsa f_{k\kp }^\bsa + \lan c_{k\bsa}^\dag f_\bsa \ran t_\sa^2 D_{k:\bsa\kp }\sum_\kpp f_{\kpp \kp}^\sa  \Big] \llan f_\sa\rran, \slabel{eq:f4c05}\\
\big[\wa_{:k}-\Sigma_{1\bsa}(\wa_{:k})\big]\langle\langle n_{\sa}
c_{ k \sa}\rangle\rangle
&=&-\lan f^\dag_\sa c_{k\sa}\ran +\sum_\kp\Big[t_\sa f_{\kp k}^\sa +  \lan f_\sa^\dag c_{k\sa} \ran  t_\sa^2 D_{\sa:k\kp}\sum_\kpp f_{\kpp \kp}^\sa\Big]\llan f_\sa\rran\nn\\
&&+ \Sigma_{5\sa}(\wa_{:k})\llan n_\bsa c_{k\sa}\rran.\slabel{eq:f4d05}
\end{subeqnarray}
where we define $f_{\kp k}^\sa \equiv \lan c^\dag_{k\bsa} c_{{k'\sa}}\ran$,
\end{widetext}
\bea
\Sigma_{1\sa}(\wa_{:k})&=&\sum_{\kp}t_\bsa^2(\wa_{\bsa:k\kp }^{-1}+\wa_{\kp:\bsa k}^{-1}),\label{eq:SE001} \\
\widehat{\Sigma}_{2\sa}(\wa_{:k})&=& \sum_{\kp\kpp} (t_\sa^2D_{\bsa :k\kp}f_{\kpp\kp}^\sa -t_\bsa^2D_{\kp:\sa k}f_{\kp\kpp}^\bsa) \nn\\
&&+\sum_{\kp} (t_\sa^2\wa_{\bsa: k\kp}^{-1}+t_\bsa^2\wa_{\kp:\sa k}^{-1}),\label{eq:SE002} \\
\widehat{\Sigma}_{3\sa}(\wa_{k:})&=&-\sum_{\kp\kpp} (t_\sa^2D_{k:\bsa\kp}f_{\kpp\kp}^\sa+t_\bsa^2D_{k:\sa\kp}f_{\kpp\kp}^\bsa) \nn\\
&&+ \sum_{\kp}(t_\bsa^2 \wa_{k:\sa \kp}^{-1}+t_\sa^2\wa_{k:\bsa\kp }^{-1}),\label{eq:SE003} 
\eea
\bea
\Sigma_{5\sa}(\wa_{:k}) &=&\wa_{:k}\sum_{\kp} t_\sa \Big[D_{\kp:\sa k}\lan c^\dag_{\kp\sa} f_\sa\ran+D_{\sa:k \kp} \lan  f_\sa^\dag c_{\kp\sa}\ran\Big]\nn\\
&&+\sum_{\kp\kpp}t_\sa^2 (D_{ \kp:\sa k}f_{\kp\kpp}^\sa-D_{\sa:k
\kp}f_{\kpp\kp}^\sa) \label{eq:SE005}
\eea
and 
\bea D_{\alpha \beta \cdots : a b \cdots} &\equiv& -U
\wa_{\alpha \beta \cdots : a b \cdots}^{-1} \left(\wa_{\alpha
\beta \cdots : a b \cdots}\pm U\right)^{-1}\label{eq:notation2}.
\eea 
In the last equation Eq.~(\ref{eq:notation2}), the sign in
front of $ U$ is the same as the sign in front of $\va_\sa$ in
$\wa_{\alpha \beta \cdots : a b \cdots}$ . Thus we have for
instance: $D_{\sa:kk'} \equiv -U \wa_{\sa:kk'}^{-1}
\left(\wa_{\sa:kk'}+ U\right)^{-1}=-U
\left(\wa+\varepsilon_{\sa}-\varepsilon_{k}-\varepsilon_{k'}\right)^{-1}
\left(\wa+\varepsilon_{\sa}-\varepsilon_{k}-\varepsilon_{k'}+
U\right)^{-1}$. Notice that we keep heuristically a
fourth-order term on the right-hand side of Eq.~(\ref{eq:f4d05}) --- explicitly the term $ \lan f_\sa^\dag c_{k\sa} \ran
\sum_{\kp\kpp}( t_\sa^2 D_{\sa:k\kp} f_{\kpp \kp}^\sa)\llan
f_\sa\rran\nn$ --- in order to respect the unitarity condition typical
of a Fermi liquid $\text{Im} [\mathcal{G}_\sa^r]^{-1} = \Gamma_\sa$. This is explained in detail in
Sec.\ref{sec:eq}. We emphasize that in principle this term shall
be recovered by properly including fourth-order contributions
generated through integral coupling of the equations of motion
(\ref{app:eq:f403}).

Combining Eqs.~(\ref{eq:f3}, \ref{eq:f4},
\ref{eq:f405}) yields the following expression for the Green function in the dot
\bea
\mathcal{G}^{r}_\sa(\wa) &=&   \frac{u_{2\sa}(\wa)-\lan n_\bsa\ran  + \Pi_\sa(\wa) }{u_{1\sa}(\wa) u_{2\sa}(\wa) - \Xi_\sa(\wa)}   , \label{eq:GF_01} 
\eea
\begin{widetext}
where we define the functions
\bea
u_{1\sa}(\wa) &=& \wa_{:\sa}-\Sigma^0_\sa(\wa) , \label{eq:u_w1} \\
u_{2\sa}(\wa) &=&  - \frac{1}{U}\left[\wa_{:\sa}-U-\sum_k\left(
\frac{t^2_\sa}{\wa_{:k}-\Sigma_{6\sa}(\wa_{:k})}+\frac{t^2_\bsa}{\wa_{\bsa:k\sa}-\widehat{\Sigma}_{2\sa}(\wa_{:k})}-
\frac{t^2_\bsa}{-\wa_{k:\sa\bsa}+U+\widehat{\Sigma}_{3\sa}(\wa_{k:})}\right)\right], \\
\label{eq:u_w2}
\Sigma_{6\sa}(\wa_{:k}) &=& \Sigma_{1\sa}(\wa_{:k})+\frac{\Sigma_{5\bsa}(\wa_{:k})\Sigma_{5\sa}(\wa_{:k})}{\wa_{:k}-\Sigma_{1\bsa}(\wa_{:k})}~,
\label{eq:sigma6} \\
\Pi_\sa(\wa) &=& -\sum_k\frac{t_\bsa\langle f_\bsa^{\dag} c_{k\bsa}\rangle}{\wa_{\bsa:k\sa}-\widehat{\Sigma}_{2\sa}(\wa_{:k})}-\sum_k\frac{t_\bsa\langle c_{k\bsa}^{\dag} f_\bsa\rangle}{-\wa_{k:\sa\bsa}+U+\widehat{\Sigma}_{3\sa}(\wa_{k:})}+\sum_k  \frac{t_{\sa}\Sigma_{5\bsa}(\wa_{:k})\lan f^\dag_\sa c_{k\sa}\ran}{[\wa_{:k}-\Sigma_{6\sa}(\wa_{:k})][\wa_{:k}-\Sigma_{1\bsa}(\wa_{:k})]} , \label{eq:Pi} \\
\Xi_\sa(\wa) &=& -\sum_{k\kp}\frac{\left[t_\bsa^2 f_{\kp
k}^\bsa +t_\bsa\lan f_\bsa^\dag c_{k\bsa} \ran  t^2_\sa D_{\bsa:k
\kp}\sum_{\kpp} f_{\kpp \kp}^\sa \right]}
{\wa_{\bsa:k\sa}-\widehat{\Sigma}_{2\sa}(\wa_{:k})} 
+ \sum_{k\kp}\frac{\left[ t_\bsa^2 f_{k\kp
}^\bsa + t_\bsa\lan c_{k\bsa}^\dag f_\bsa \ran t_\sa^2
D_{k:\bsa\kp}\sum_\kpp f_{\kpp \kp}^\sa \right]}
{-\wa_{k:\sa\bsa}+U+\widehat{\Sigma}_{3\sa}(\wa_{k:})} \nn \\
&& + \sum_{k\kp}  \frac{ \left[t_{\sa}^2  f_{\kp
k}^\sa +t_{\sa} \lan f_{\sa}^\dag c_{k\sa} \ran t_\sa^2
D_{\sa:k\kp}\sum_\kpp f_{\kpp \kp}^\sa\right]
\Sigma_{5\bsa}(\wa_{:k})}{[\wa_{:k}-\Sigma_{6\sa}(\wa_{:k})][\wa_{:k}-\Sigma_{1\bsa}(\wa_{:k})]} .
\label{eq:Xi}
\eea
\end{widetext}
In order to close the problem, one needs to append to Eq.(\ref{eq:GF_01}) the closure equations, which enables one to determine the expectation values showing up in the expression of $\mathcal{G}^{r}_\sa(\wa)$. The calculations of the expectation values are presented in \ref{App:TA1}, while the full self-consistent treatment is explained in Sec.\ref{sec:thermal average}. $\mathcal{G}^{r}_\sa(\wa)$ given by Eq.  (\ref{eq:GF_01}) respects charge conjugation symmetry, as proved in \ref{App:CCS}.

\section{Analytical results\label{sec:analytic}}

\begin{table*}
    \centering
        \begin{tabular}{|c|c|c|c|c|}
     \hline & & & &  \\
             & $\quad \gamma_{1\sa}^{(2)} \quad$ & $\quad \gamma_{2\sa}^{(2)} \quad$ & $\quad \gamma_{3\sa}^{(2)} \quad$ & $\quad \gamma_{5\sa}^{(2)} \quad$  \\
             & & & & \\
             \hline & & & &  \\
             Empty dot & $\Gamma$ & $\Gamma$ & $\Gamma$ & 0   \\
             $\left(\va_\sa-\mu_\alpha \gg \Gamma\right)$ & & & &  \\
             \hline & & & &  \\
             Kondo regime & $\Gamma$ & 0 & $2\Gamma$ & $\Gamma$  \\
             $\left(\va_\sa+U-\mu_\alpha,~\mu_\alpha-\va_\sa \gg \Gamma\right)$ & & & &  \\
             \hline & & & &  \\
             Doubly-occupied dot & $\Gamma$ & $\Gamma$ & $\Gamma$ & 0  \\
             $\left(\mu_\alpha-\va_\sa-U \gg \Gamma\right)$ & & & &  \\
             \hline & & & &  \\
             Mixed valence regime & $\Gamma$ & $\Gamma$ & $\Gamma$ & 0  \\
             $\left(  {\rm Min}\{\va_\sa-\mu_\alpha,\mu_\alpha-\va_\sa-U\} \approx
\Gamma\right)$ & & & &  \\
             \hline
        \end{tabular}
    \caption {\protect\small {
    Transition rates $\gamma_i^{(2)} = -{\rm Im}\Sigma_{i}$ at the second order in $t_\sa$ and at zero temperature, for the different regimes of the Anderson model obtained by the EOM approach. Notice that, in the Kondo regime, $\gamma_{2}^{(2)}=0$ yields low-energy logarithmic divergence of the self-energy of the dot Green function, responsible for the Kondo effect. In the latter regime, $\gamma_{5\sa}^{(2)} \neq 0$, which brings on an additional divergence arising from Eq.~(\ref{eq:log_1}).   
    }}
    \label{tab:gam_sec_order}
\end{table*}

In this section, we discuss some aspects of the behavior of the
system in and out of equilibrium for the different regimes of the
Anderson model, as can be derived from the results obtained in the previous
section. Special care is given to the singly-occupied dot regime
where many-body effects can give rise to Kondo physics. We analyze
in detail the nonequilibrium situation in the latter regime, and
show how the EOM method provides a powerful frame to describe the
decoherence effects induced when a bias voltage is applied to the
leads.

\subsection{Renormalization effects and transition rates}
In the presence of the Coulomb interaction $U$ and the dot-lead tunneling
coupling $\Gamma_{\alpha\sa}$, the bare parameters of the
Anderson model get renormalized according to (for zero
temperature) \small
\begin{subeqnarray}
\label{Renorm_dotlevel}
\va^*_\sa \simeq \va_\sa - \displaystyle \sum_\alpha \frac{\Gamma_{\alpha\bsa}}{\pi}{\rm ln}\left(\frac{\left|\va_\bsa^*-\mu_\alpha \right|}{{\rm Min}\{W,\left|\va_\bsa^*+U^*-\mu_\alpha \right|\}}\right)  , \quad \\
U^*_\sa \simeq U + \displaystyle \sum_{\alpha\sa}
\frac{\Gamma_{\alpha\sa}}{\pi}{\rm
ln}\left(\frac{\left|\va_\sa^*-\mu_\alpha \right|}{{\rm
Min}\{W,\left|\va_\sa^*+U^*-\mu_\alpha \right|\}}\right) . \quad
\end{subeqnarray}
\normalsize 
The above results are obtained from
Eq.(\ref{eq:GF_01}) up to second order in $t_{\sigma}$ and by taking
$f_{kk'}^{\sigma}=f^\alpha_{F}(\va_{k})\delta_{kk'}$. In
the mixed valence regime $\left({\rm Min}\{\va_\sa-\mu_\alpha,\mu_\alpha-\va_\sa-U\} \approx
\Gamma\right)$, the renormalization of the bare level energy is
consistent with the prediction of the scaling theory
\cite{HaldanePRL78,Hewson} as pointed
out in the previous EOM studies\cite{Lacroix1981}. As expected,
the renormalization effects are small around the particle-hole
symmetric case ($\varepsilon_{\sigma}=-U/2$). In the large
$U/|\varepsilon_{\sigma}|$ limit, the renormalization effects are
very important, as it is the case for quantum dots
coupled to ferromagnetic leads\cite{Martinek2003,Yasuhiro2005}.\

Interestingly, these renormalizations are consistent with the
shift of the pole of the Green functions $\langle\langle
f_\bsa^{\dag}c_{k\bsa}f_{\sa}\rangle\rangle$ and $\langle\langle
c_{k\bsa}^{\dag}f_\bsa f_{\sa}\rangle\rangle$. For instance from
Eq. (\ref{eq:f4b05}), the pole of $\langle\langle f_\bsa^{\dag}c_{k\bsa}f_{\sa}\rangle\rangle$
(with respect to $\wa_{:k}$) is shifted to, 
\bea
\va_{\sa}-\va_{\bsa}+ \text{ Re
}\widehat{\Sigma}_{2\sa}(\va_{\sa}^*-\va_{\bsa}^*) = \va_{\sa}^*-\va_{\bsa}^*. \nn 
\eea 
The shift of these poles can have important consequences on the splitting of
the Kondo resonance peak when a magnetic field is applied. These
corrections are neglected in the Lacroix approximation.\

The imaginary part of the corresponding self-energies evaluated at
the pole of the Green functions (e.g. at
$\wa_{:k}=\va^*_\sa-\va^*_\bsa$ for $\langle\langle
f_\bsa^{\dag}c_{k\bsa}f_{\sa}\rangle\rangle$) defines the transition rate from the state
$f_\sa^{\dag}|GS\rangle)$ to the excited state $f_\sa^{\dag}c_{k\bsa}^{\dag}f_\bsa|GS\rangle$, where
the ground state is denoted by $|GS\rangle$. Within second order
in $t_{\sigma}$ and taking into account the renormalization of the
dot level energies, the transition rates are given by
\begin{subeqnarray}
\gamma_{1\sa}^{(2)}&=& -{\rm Im} \Sigma_{1\sa}(0) = 2\Gamma_{\bsa} , \\
\gamma_{2\sa}^{(2)}&=&-{\rm Im}\widehat{\Sigma}_{2\sa}(\va^*_\sa-\va^*_\bsa)\nn\\
&=&\sum_{\alpha=L,R}\sum_{\sa}\Gamma_{\alpha \sa}[1- f^\alpha_F(\va^*_\sa)+f^\alpha_F(\va_\sa^*+U^*)], \quad \; \label{eq:lifetime01}\\
\gamma_{3\sa}^{(2)}&=&-{\rm Im}\widehat{\Sigma}_{3\sa}(\va^*_\sa+\va^*_\bsa +U^*)\nn\\
&=&\sum_{\alpha=L,R}\sum_{\sa}\Gamma_{\alpha \sa}[1+f^\alpha_F(\va^*_\sa)-f^\alpha_F(\va^*_\sa+U^*)],\label{eq:lifetime02}\\
\gamma_{5\sa}^{(2)}&=&-{\rm Im}\Sigma_{5\sa}(0)\nn\\
&=&2\sum_{\alpha=L,R}\Gamma_{\alpha\sa}\left[f^\alpha_F(\va^*_{\sa})-f^\alpha_F(\va^*_{\sa}+U^*)\right] ,\label{eq:lifetime03}
\end{subeqnarray}
where $\Gamma_{\sa}=\Gamma_{L\sa}+\Gamma_{R\sa}$ using the
notation defined in Sec.\ref{secEOM}. The values of these
second-order transition rates in the case of spin-independent tunneling ($\Gamma_\uparrow=\Gamma_\downarrow=\Gamma/2$) are reported in Table
\ref{tab:gam_sec_order} for the different regimes of the Anderson
model at zero temperature. One can note that, in the wide-band limit, the value of $\gamma_{1\sa}^{(2)}$ does not
depend on the occupancy in the dot. In contrast, the other transition
rates take different values depending on the regimes considered.
One can distinguish four regimes:\

\begin{enumerate}
    \item[a)] In both the empty and doubly-occupied dot regimes, $\gamma^{(2)}_{5\sa} \simeq 0$ and $\gamma_{2\sa}^{(2)}$,
$\gamma_{3\sa}^{(2)} \simeq \Gamma$.
As $\gamma_{5\sa}^{(2)}=0$ in these two regimes, the third term of $\Xi_{\sigma}(\wa)$ and $\Pi_{\sigma}(\wa)$ vanishes (cf. Eqs. (\ref{eq:Pi},\ref{eq:Xi})). On the other side, the finite values of $\gamma_{2\sa}^{(2)}$ and $\gamma_{3\sa}^{(2)}$ provide a cut-off
to the integrals involved in the calculation of the remaining
terms, thereby preventing them from diverging at low energy. As a result, the electron density of states in the dot does not show any resonance peak but only two
broad peaks located at the positions of the renormalized dot level
energies.
    \item[b)]  In the mixed valence regime (take for instance $\va_\sa-\mu_\alpha \approx \Gamma$), the renormalization effects
push the dot level energies above the
chemical potential, hence the transition rates are
identical to those found in the two regimes of a). Our numerical results for the
density of states are in better agreement with the
exact numerical renormalization group result than those found in
the Lacroix approximation or the non-crossing approximation, for
which a spurious peak may appear at the Fermi level, as it has been
shown in Ref.\onlinecite{Monreal2005}.
    \item[c)] The singly-occupied dot (Kondo) regime is the most
    interesting since one of the transition rates $\gamma_{2\sa}^{(2)}$ vanishes.
    This gives rise to a logarithmical divergence at low energy of the integral
    involved in the calculation of the first term of $\Xi_{\sigma}(\wa)$ and $\Pi_{\sigma}(\wa)$ in Eqs. (\ref{eq:Pi},\ref{eq:Xi}). Another divergence comes from the calculation of the third term of $\Xi_{\sigma}(\wa)$ and $\Pi_{\sigma}(\wa)$, which no longer vanishes as $\gamma_{5\sa}^{(2)}$ is now finite.
    The integrand of those terms has a structure like
    \bea
\frac{\Sigma_{5\bsa}(\wa_{:k})}{[\wa_{:k}-\Sigma_{6\sa}(\wa_{:k})][\wa_{:k}-\Sigma_{1\bsa}(\wa_{:k})]}  \nn \\
\simeq
-\frac{\Gamma_\bsa}{\Gamma}\left(\frac{1}{\wa_{:k}}-\frac{1}{\wa_{:k}+2i\Gamma}\right)
+ \mathcal{O}\left(\frac{\Gamma \wa_{:k}}{U} \right) , \label{eq:log_1} 
\eea
These terms have
two poles at $\wa_{:k}=0+i\gamma_{5a\sa}^{(2)}$ and
$\wa_{:k}=0+i\gamma_{5b\sa}^{(2)}$ with
$\gamma_{5a\sa}^{(2)} = \gamma_{1\bsa}^{(2)}-\gamma_{5\sa}^{(2)}
=0$ and $\gamma^{(2)}_{5b\sa} =
\gamma_{1\bsa}^{(2)}+\gamma_{5\sa}^{(2)} =2\Gamma$, respectively. The values of
the imaginary part of these two poles are reported in Table \ref{tab:gam_sec_order} for
the Kondo regime. Since $\gamma_{5a\sa}^{(2)}=0$, the first term
in Eq.~(\ref{eq:log_1}) gives rise to an
additional logarithmical divergent self-energy term at low energy.\
The presence of these two logarithmical divergences mentioned above is
solely responsible for the formation of the Kondo resonance peak in the
electron density of states in the dot. In Sec.\ref{sec: Tk}, we will analytically estimate the Kondo temperature from the consequences
of these divergent self-energy terms.

\end{enumerate}

\subsection{Case of the Kondo regime}

The Kondo regime is particularly interesting because some
logarithmical divergent terms (Kondo singularities) survive even
after introducing second-order self-energy corrections $\Sigma_{i\sa}^{(2)} (i=1,2,3,5)$, as discussed before. We focus in more detail on this regime and
show how fourth-order corrections in $t_\sigma$ smear the
Kondo singularities when the system is driven out of equilibrium.

\subsubsection{Compact expression for the electron Green function in the dot}

To facilitate the understanding of Eq.~(\ref{eq:GF_01}), it is
instructive to put the expression of $\mathcal{G}^{r}_\sa(\wa)$ in the Kondo regime in a more compact way in order to better identify
the terms bringing about Kondo singularities. After integrating
over $k$ and using the closure
equations for the expectation values (see \ref{App:TA1}), we can express the functions $\Xi_{\sigma}(\wa)$ and $\Pi_{\sigma}(\wa)$ appearing in $\mathcal{G}^{r}_\sa(\wa)$  as 
\bea
\Xi_{\sigma}(\wa) &=& -\mathcal{F}^+ \circ Q_\sa(\wa) - i \Gamma_\sa \mathcal{F}^- \circ P_\sa(\wa)  \label{shortXi} , \\
\Pi_{\sigma}(\wa) &=& \mathcal{F}^- \circ  P_\sa(\wa) \label{shortPi} .
\eea
where the functional
$\mathcal{F}^\pm$ acting on any function $X_\sa(\wa)$ is defined
as 
\bea
\mathcal{F}^\pm \circ X_\sa(\wa) &=&  -X_\bsa\Big(\wa_{\bsa^*:\sa^*} + i\gamma_{2\sa}^{(4)}\Big) \nn \\
&& \pm X_\bsa\Big(-\wa_{:\sa^*\bsa^*}+U^*-2i\Gamma\Big) \nn \\
&& + \frac{\Gamma_\bsa}{\Gamma}  \left[- X_\sa\Big(\wa +
i\gamma_{5a\sa}^{(4)}\Big) +  X_\sa\Big(\wa+ 2 i \Gamma \Big)
\right] , \nn \\ \label{eq:GF_002} 
\eea 
where `$\sa^*$' in $\wa_{\bsa^*:\sa^*}$ and $\wa_{:\sa^*\bsa^*}$ indicates that the dot level energies are renormalized.\

The derivation of $P_\sa(\wa)$ and $Q_\sa(\wa)$ is given in
\ref{App:TA1}, while the full self-consistent treatment is
discussed in Sec.\ref{sec:thermal average}. We report here the
result obtained for $P_\sa(\wa)$ and $Q_\sa(\wa)$
\begin{align}
P_\sa(\wa)&=\sum_{\alpha=L,R}\frac{\Gamma_{\alpha\sa}}{\pi }\int d\va \frac{f^\alpha_F(\va)\mathcal{G}^a_\sa(\va)}{\wa-\va+i\delta},  \label{eq:taP0}\\
Q_\sa(\wa)&=\sum_{\alpha=L,R}\frac{\Gamma_{\alpha\sa}}{\pi }\int
d\va
\frac{f^\alpha_F(\va)[1+i\Gamma_\sa\mathcal{G}_\sa^{a}(\va)]}{\wa-\va+i\delta} ,
 \label{eq:taQ0}
\end{align}
where $\mathcal{G}^a_\sa(\va)$ is the advanced dot Green function.\

One can see from Eqs.~(\ref{eq:GF_01},\ref{shortXi},\ref{shortPi},\ref{eq:GF_002}) that the expression of $\mathcal{G}^{r}_\sa(\wa)$ contains four terms
$Q_\bsa(\wa_{\bsa^*:\sa^*}),~P_\bsa(\wa_{\bsa^*:\sa^*})$,
$Q_\sa(\wa)$ and $P_\sa(\wa)$ which give rise to low-energy Kondo
singularities when only second-order transition rates are considered.
We will see in the next section how the nonequilibrium situation
cures these divergences by introducing finite transition rates coming
from fourth-order contributions in $t_\sa$ which provide a cut-off
energy to the divergent integral terms.

\subsubsection{Decoherence rates induced out of equilibrium \label{sec:4order_lifetimes}}

We calculate explicitly the fourth-order transition rates (decoherence
rates) by expanding the equations of motion to sixth order in $t_\sigma$,
followed by the usual truncation. The derivation is long but
straightforward and we present only the results for the fourth-order decoherence rates in the Kondo regime, namely, $\gamma_{2\sa}^{(4)}$ and
$\gamma_{5a\sa}^{(4)}$
\begin{widetext}
\begin{subeqnarray}
\label{eq:tau25}
\gamma_{2\sa}^{(4)}&=&\sum_{\alpha,\beta=L,R}\sum_{\sa,\sa^\prime}\frac{\Gamma_{\alpha\sa} \Gamma_{\beta\sa^\prime}}{\pi} \int d\va (1-f_F^\alpha(\va ))f_F^\beta(\va-\va_\sa+\va_{\sa^\prime}) \mathcal{P}[D_\sa(\va)^2],\slabel{eq:tau2}\\
\gamma_{5a\sa}^{(4)}&=&\sum_{\alpha,\beta=L,R}\sum_{\sa,\sa^\prime\atop
\sa\not=\sa^\prime}\frac{2\Gamma_{\alpha\sa}
\Gamma_{\beta\sa^\prime}}{\pi} \int d\va (1-f_F^\alpha(\va
))f_F^\beta(\va-\va_\sa+\va_{\sa^\prime})
\mathcal{P}[D_\sa(\va)^2],\slabel{eq:tau5}
\end{subeqnarray}
where
\bea
D_\sa(\va)= \frac{1}{\va-\va_\sa+i\delta}-\frac{1}{\va-\va_\sa-U+i\delta}
.
\label{eq:Dfunc}
\eea
In the limit $V = \left|\mu_L-\mu_R\right| \ll {\rm Min}\{\left|\va_d-\mu_{eq}\right|, ~\va_d-\mu_{eq}+U\}$ (with $\mu_{eq}=\left(\mu_L-\mu_R\right)/2$), and at zero
temperature
\begin{subeqnarray}
\label{eq:tau67}
\gamma_{2\sa}^{(4)} 
&\approx& \sum_{\alpha,\beta=L,R}\sum_{\sa^{\prime},\sa^{\prime\prime}}\frac{\Gamma_{\alpha\sa^{\prime}} \Gamma_{\beta\sa^{\prime\prime}}}{\pi}(\mu_\beta-\mu_\alpha+\va_ {\sa^{\prime}}-\va_{\sa^{\prime\prime}})\Theta(\mu_\beta-\mu_\alpha+\va_ {\sa^{\prime}}-\va_{\sa^{\prime\prime}})D_ {\sa^{\prime}}(\mu_\alpha)D_{\sa^{\prime\prime}}(\mu_\beta)\nn\\
&=&\frac{\pi}{4}\sum_{\alpha,\beta=L,R}\sum_{\sa^{\prime},\sa^{\prime\prime}}(\mu_\beta-\mu_\alpha+\va_ {\sa^{\prime}}-\va_{\sa^{\prime\prime}})\Theta(\mu_\beta-\mu_\alpha+\va_ {\sa^{\prime}}-\va_{\sa^{\prime\prime}})\rho^0_\alpha \rho^0_\beta J_{\alpha\sa^{\prime} , \beta\sa^{\prime\prime}}J_{\beta\sa^{\prime\prime} ,\alpha\sa^{\prime}  }, \slabel{eq:tau6}\\
\gamma_{5a\sa}^{(4)} &\approx& \sum_{\alpha,\beta=L,R}\sum_{\sa^{\prime},\sa^{\prime\prime}\atop \sa^{\prime}\not=\sa^{\prime\prime}}\frac{2\Gamma_{\alpha\sa^{\prime}} \Gamma_{\beta\sa^{\prime\prime}}}{\pi}(\mu_\beta-\mu_\alpha+\va_ {\sa^{\prime}}-\va_{\sa^{\prime\prime}})\Theta(\mu_\beta-\mu_\alpha+\va_ {\sa^{\prime}}-\va_{\sa^{\prime\prime}})D_ {\sa^{\prime}}(\mu_\alpha)D_{\sa^{\prime\prime}}(\mu_\beta)\nn\\
&=&\frac{\pi}{2}\sum_{\alpha,\beta=L,R}\sum_{\sa^{\prime},\sa^{\prime\prime}\atop
\sa^{\prime}\not=\sa^{\prime\prime}}(\mu_\beta-\mu_\alpha+\va_ {\sa^{\prime}}-\va_{\sa^{\prime\prime}})\Theta(\mu_\beta-\mu_\alpha+\va_ {\sa^{\prime}}-\va_{\sa^{\prime\prime}})\rho^0_\alpha
\rho^0_\beta J_{\alpha\sa^{\prime},
\beta\sa^{\prime\prime}}J_{\beta\sa^{\prime\prime},\alpha\sa^{\prime}  },\slabel{eq:tau7}
\end{subeqnarray}
\end{widetext}
where $\Theta(x)$ is the Heaviside step function and $\mathcal{P}$
denotes the principal value of a function. Eqs.~(\ref{eq:tau67}) are expressed in terms of the Kondo exchange
coupling\cite{Hewson} $J_{\alpha\sa , \beta\sa^\prime }\equiv 2
t_{\alpha\sa } t_{\beta\sa^\prime } D_\sa(\mu_\alpha)$. In the absence of magnetic field, both decoherence rates are equal $\gamma_{2\sa}^{(4)}=\gamma_{5a\sa}^{(4)}=\gamma^{(4)}$.

The expressions of these two decoherence rates are the main result
of this section. Although for $ \gamma_{5a\sa}^{(4)}$, summation
is only over opposite spins, both of them involve at least one
spin-flip process. At zero temperature, these decoherence rates
are finite as soon as a bias voltage and/or a Zeeman splitting is introduced. The finite values of these decoherence rates
provide a cut-off to the divergent integral terms of the Green function and smear the Kondo singularities. Note that
$\gamma_{2\sa}^{(4)}$ is slightly different from the heuristical
result of Ref.\onlinecite{MWL1993} obtained from the
Fermi golden rule, because here {\it both} spins contribute to the
rate. Our overall result for the decoherence effect is
consistent with those found using a real-time diagrammatic
technique\cite{Yasuhiro2005} and the non-crossing approximation
\cite{WingreenPRB94}, although in the latter case the decoherence
rate was not calculated explicitly.

\subsubsection{Kondo temperature\label{sec: Tk}}

At equilibrium and at zero temperature, the Kondo scale $T_K$ ($k_B=1$) can be
roughly estimated from the zero of the real part of the
denominator of $\mathcal{G}^{r}_\sa(\wa)$ in Eq.~(\ref{eq:GF_01}) located near the chemical
potential\cite{Hewson}. Considering the case of zero magnetic field and spin-independent
couplings $\Gamma_\sa=\Gamma_\bsa=\Gamma/2$ in the
wide-band limit, $T_K$ reads 
\bea 
T_{K} \simeq \left[2\Gamma (2\va_0+U)^2+ 8\Gamma^3\right]^{1/3} {\rm exp}\left\{\frac{4\pi \va_0(\va_0+U)}{3\Gamma U}\right\} , \label{eq:TkMF} 
\eea
where $\va_0=\va_d-\mu_{eq}$. $T_K$ is independent of $W$, as expected since the
high-energy scale is now regulated by $U$.

We now compare our result for $T_K$ in Eq.(\ref{eq:TkMF}) with that obtained within the Lacroix approximation
$\displaystyle[2\va_0+U]{\rm exp}\left[2\pi \va_0(\va_0+U)/\Gamma
U\right]$.

First, the Lacroix result for $T_K$ is improved by an
exponential factor 4/3, in better agreement with Haldane's
prediction \cite{HaldanePRL78} $\displaystyle
(U\Gamma/4)^{1/2}{\rm exp}\left[\pi \va_0(\va_0+U)/\Gamma
U\right]$. This is due to the presence of an additional
logarithmical divergent term in the self-energy given by
$\Xi_{\sigma}(\va)$. To our best knowledge, this contribution, coming from a fourth-order self-energy, was first found
by Dworin\cite{Dworin1967} and was attributed to a finite lifetime
mechanism of the localized electron. It was lately reproduced
\cite{Monreal2005} in the infinite $U$ limit.

Secondly, at the particle-hole symmetric point $(2\va_0+U=0)$, the
proposed approximation cures the aforementioned pathology of the
Lacroix approximation for which $T_K$ vanishes, as
will be further explained in Sec.\ref{sec:eq}. The expression of
$T_K$ at that point is given by \bea T_K \simeq
2\Gamma ~{\rm exp}\left[\frac{2\pi \va_0}{3\Gamma}\right]. \eea

\section{Numerical Results\label{secResult}}

We present our numerical results in and out of equilibrium,
and discuss the evolution of the density of states as well as
transport quantities. 
We consider a quantum dot connected symmetrically to
the two leads with spin-independent tunneling couplings
($\Gamma_{L\sigma}=\Gamma_{R\sigma}=\Gamma_{L\bsa}=\Gamma_{R\bsa}=\Gamma/4$), and take a large ratio $U/\Gamma < W/\Gamma=20$ in order to be in the wide-band limit. For illustrative purposes, we choose to present the results at the particle-hole symmetric point
($\varepsilon_{d}=-U/2$), which turns out to be particularly well
described by our method, in contrast with the other EOM approaches developed so far. Finally, we
limit the study to the case of zero magnetic field in order to
concentrate on the nonequilibrium effects brought by the
application of a bias voltage. 
The Green function $\mathcal{G}^{r}_\sa(\wa)$ given by Eq.~(\ref{eq:GF_01}) is solved in a fully self-consistent
way (cf. Sec.\ref{sec:thermal average}). 
In the Kondo regime, an important energy scale is
provided by the Kondo temperature which needs to be properly
defined. We will not use the approximate expression for $T_K$ given
by Eq.(\ref{eq:TkMF}) but rather calibrate it
numerically from the temperature dependence of the zero-bias
conductance 
\bea 
\left.\frac{dI}{dV}\right|_{T=T_K ; V=0} = \frac{1}{2} G_0 ,
\eea 
where $G_0(=2e^2/h)$ is the zero-bias conductance at zero temperature.

\subsection{Self-consistency\label{sec:thermal average}}

The dot Green function given by
Eq.~(\ref{eq:GF_01}) shows an explicit dependence on the
expectation values $\langle f_\sa^{\dag} c_{k\sa} \rangle$,
$\langle c_{k\sa}^{\dag} c_{k'\sa} \rangle$ (denoted by $f_{\kp
k}^\sa$ previously) and $\lan n_\sa\ran=\langle f_\sa^{\dag}
f_{\sa}\rangle$. What matters then is to compute these expectation
values in order to properly define the self-consistency scheme. In
general (in both equilibrium and nonequilibrium situations), the
expectation values (as for instance $\langle f_\sa^{\dag} c_{k\sa}
\rangle$) can be expressed in terms of the related lesser Green
function \bea \lan f^\dag_\sa c_\ka \ran \equiv -i \int
\frac{d\wa}{2\pi}\mathcal{G}^<_{\ka,\sa}(\wa). \eea

In equilibrium, the relationship
$\mathcal{G}^<_{\ka,\sa}(\wa)=-f_F(\wa)
[\mathcal{G}^r_{\ka,\sa}(\wa)-\mathcal{G}^a_{\ka,\sa}(\wa)]$
holds, relating the lesser to the retarded and advanced Green
functions, $\mathcal{G}^r_{\ka,\sa}(\wa)$ and
$\mathcal{G}^a_{\ka,\sa}(\wa)$ respectively. The expectation value
is then given by 
\bea 
\langle f_\sa^{\dag} c_{k\sa} \rangle
=-\frac{1}{\pi}\int d\wa f_F(\wa){\rm
Im}\mathcal{G}^r_{\ka,\sa}(\wa)\label{eq:SF01} .
\eea 
This relationship is nothing else but the spectral theorem which expresses the
expectation value in terms of a functional of the corresponding
retarded Green function. As a result, in
equilibrium, Eq.~(\ref{eq:GF_01}) ends up being an integral
equation with respect to $\mathcal{G}^r_{\sigma}(\wa)$ that can be
solved self-consistently.

However, out of equilibrium, the above relationship between the
different Green functions no longer holds, and one cannot compute
the expectation values from the spectral theorem. An alternative
is to work within the Keldysh formalism. The details of the calculations of the
expectation values $\langle f_\sa^{\dag} c_{k\sa} \rangle$ and
$\langle c_{k\sa}^{\dag} c_{k'\sa} \rangle$ within our EOM
approach, and of the related integrals $P_\sa(\wa)$ and
$Q_\sa(\wa)$ through which these expectation values contribute to
Eq.~(\ref{eq:GF_01}), are presented in \ref{App:TA1} (cf.
Eqs.~\ref{eq:taP}-\ref{eq:taQ}). In the wide-band limit, it turns out that even out of equilibrium, the
integrals $P_\sa(\wa)$ and $Q_\sa(\wa)$ keep the same structure as in equilibrium, and depend only on the
retarded Green function without requiring any knowledge of the
lesser Green function.

As far as the occupation number in the dot $\lan n_\sa\ran=\lan
f^\dag_\sa f_\sa\ran$ is concerned, the calculation is rather more
complicated out of equilibrium since the
simplification which takes place before for the calculation of
$P_\sa(\wa)$ and $Q_\sa(\wa)$ does not occur, and one needs to
know the lesser Green function $\mathcal{G}^<_{\sa}(\wa)$ in order
to derive $\lan n_\sa\ran$ by the use of 
\bd \lan n_\sa\ran \equiv
-i \int
\frac{d\wa}{2\pi}\mathcal{G}^<_{\sa}(\wa) . \label{eq:def_ocuN} 
\ed
To find $\mathcal{G}^<_{\sa}(\wa)$, we use the Dyson
equation written in the Keldysh formalism
$\mathcal{G}_\sa^<(\wa)=\mathcal{G}_\sa^r(\wa) \Sigma_\sa^<(\wa)
\mathcal{G}_\sa^a(\wa)$ and express the lesser self-energy
$\Sigma_\sa^<(\wa)$ via the Ng ansatz\cite{Ng96}
\[
\Sigma_\sa^<(\wa)=-2i\sum_{\alpha=L,R}\frac{\Gamma_{\alpha
\sa}}{\Gamma_\sa}f_F^\alpha(\wa){\rm Im}\Sigma_\sa^r(\wa) ,
\]
where $\Sigma_\sa^r(\wa) \equiv \wa - \va_\sa - [\mathcal{G}^r_\sa (\wa)]^{-1}$ is the retarded self-energy. This ansatz
is based on an extrapolation from both the non-interacting limit out of equilibrium and the interacting limit in
equilibrium. Thanks to this ansatz, the
calculation of $\lan n_\sa\ran$ can be performed from the
knowledge of $\mathcal{G}^r_{\sigma}(\wa)$ only. Let us also mention that many results can be obtained at the particle-hole symmetric point (also out of equilibrium), where the occupation number is identically $1/2$.

Therefore, all the expectation values
relevant to the calculations can be expressed in terms of $\mathcal{G}^{r}_\sa(\wa)$, and the
self-consistent scheme is straightforward. Eq.~(\ref{eq:GF_01}) ends
up being again a complex integral equation with respect to
$\mathcal{G}^r_{\sigma}(\wa)$, exactly as in the equilibrium
situation except that now the different chemical potentials of the
two leads have to be entered explicitly. We emphasize that this
constitutes a huge simplification in the technique that renders
the approach developed in Sec.\ref{secEOM} tractable in a self-consistent scheme even out of equilibrium.

\begin{figure}[t]
\epsfig{figure=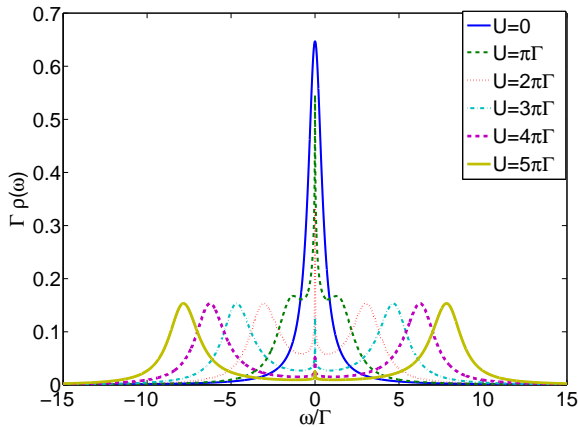,clip=,width=3.4 in}
\caption{{\protect\small {(Color online) Equilibrium density of states in the particle-hole symmetric case at $T/\Gamma=10^{-3}$
for different values of the parameter $U$ (the chemical potential of the
lead $\mu_{eq}$ is taken equal to $0$). The density of states for
large $U$ shows a three-peak structure with two broad side peaks
and a narrow Kondo resonance peak centered at the Fermi level.}}}
\label{fig:DOSMF_PH}
\end{figure}

\begin{figure}[t]
\begin{center}
\epsfig{figure=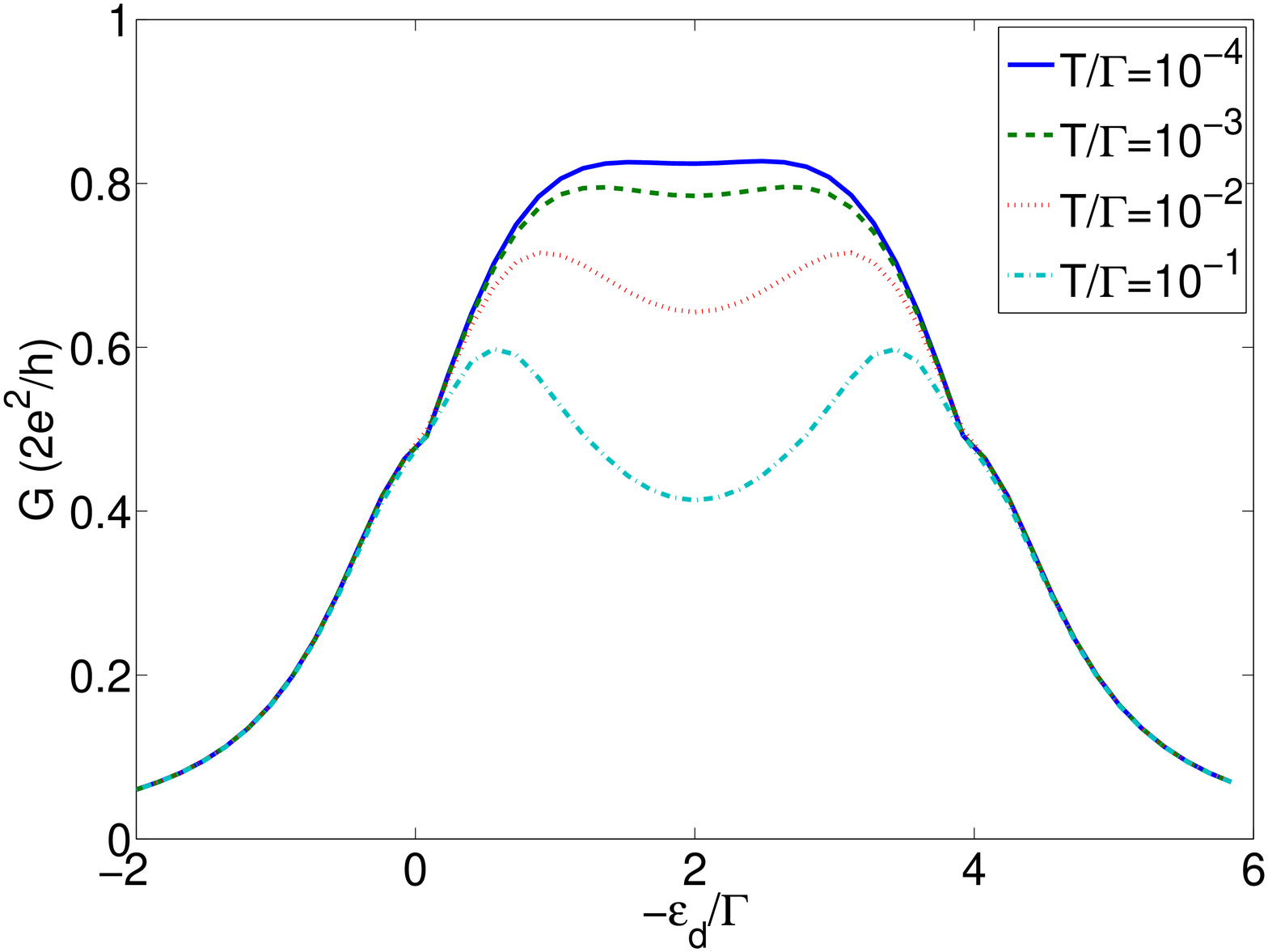,clip=,width=3.4 in}
\end{center}
\vspace{-0.3cm} \caption{Linear conductance as a function of dot level energy $\va_d$, for
$U/\Gamma=4$ and at different temperatures. When the temperature is lowered, the conductance is
enhanced in the singly-occupied regime $-\va_d/\Gamma \in [0,4]$,
and eventually reaches the maximum conductance $2e^2/h$ for a
single channel at zero temperature. The conductance does not reach this limit here because of the numerical accuracy of the self-consistent treatment.} \label{fig:G_vs_ed}
\end{figure}

\subsection{In equilibrium\label{sec:eq}}
We compute the density of states in the dot
$\rho_\sa(\wa)=-1/\pi{\rm Im}\mathcal{G}_{\sa}^{r}(\wa)$ at
equilibrium using our EOM approach. Fig.\ref{fig:DOSMF_PH} reports
the result for the density of states at equilibrium and
$T/\Gamma=10^{-3}$ for different values of the parameter $U$ when
the value of the Fermi level of the leads $\mu_{eq}$ is taken equal to
zero. We willingly choose to consider the particle-hole symmetric
case ($\varepsilon_{d}=-U/2$) since we know that it is a delicate
case in the sense that the EOM approaches developed so far have failed to describe it correctly. The density of states shows a three-peak
structure as soon as $U$ becomes larger than $\Gamma$, with two
broad peaks and a narrow Kondo resonance peak. The two broad peaks are centered at the renormalized energy levels; their
position, intensity and amplitude agree quantitatively with the
NRG result\cite{Costi90}. The Kondo resonance peak is pinned at
the Fermi level of the leads.

The fact that our EOM scheme correctly describes the
particle-hole symmetric case is one of the successes of the
method. This can be understood by the fact that in the previous
EOM approaches, for the Kondo regime, there is an exact
cancellation of the divergent terms
$Q_\bsa(\wa_{\bsa:\sa})-Q_\bsa(-\wa_{:\sa\bsa}+U) =
Q_\bsa(\wa)-Q_\bsa(-\wa)=0$. This feature is cured in our EOM
approach since the function $Q_\bsa(-\wa_{:\sa\bsa}+U)$ in
Eq.~(\ref{shortXi}) acquires a finite transition rate $2\Gamma$ and
is therefore smeared out. Therefore, the cancellation does not occur any
longer, and we are left with a divergence in the self-energy at the
origin of the formation of the Kondo resonance peak. Through the same argument, our approach is shown in
Sec.\ref{sec: Tk} to improve the prediction made previously by the Lacroix approximation for the Kondo temperature in the particle-hole
symmetric case.

Moreover, the density of states at the Fermi level is found to be
$\rho_\sa(\mu_{eq})=2/\pi\Gamma$ in agreement with the Fermi liquid
property at zero temperature and hence respecting the unitarity
condition. This can be explained as follows: at zero temperature,
the functions $P_\sa(\wa)$ and $Q_\sa(\wa)$ diverge logarithmically as
$\wa\rightarrow \mu_{eq}$,
\begin{subeqnarray}
P_\sa(\wa)&=&-\frac{\Gamma_\sa}{\pi}\mathcal{G}^a_\sa(\mu_{eq}){\rm ln}\left|\wa-\mu_{eq}\right|+\mathcal{O}(1),\nn\\
Q_\sa(\wa)&=&-\frac{\Gamma_\sa}{\pi}[1+i\Gamma_\sa\mathcal{G}^a_\sa(\mu_{eq})]{\rm
ln}\left|\wa-\mu_{eq}\right|+\mathcal{O}(1).\nn
\end{subeqnarray}
We find that the inverse of the imaginary part of $\mathcal{G}^{r}_\sa(\wa)$ (cf.Eqs.~(\ref{eq:GF_01},\ref{shortXi})) is ${\rm
Im}[\mathcal{G}^{r}_\sa]^{-1}(\mu_{eq}) =\Gamma/2$, as expected from the
Fermi liquid theory. In the
particle-hole symmetric case, we find ${\rm Re}\mathcal{G}^{r}_\sa(\mu_{eq}) \simeq 0$. Combining these two
results leads to $\rho_\sa(\mu_{eq})=2/\pi\Gamma$ as observed in
Fig.\ref{fig:DOSMF_PH}. Inserting the value of $\rho_\sa(\mu_{eq})$
into Eq.~(\ref{Landauer}) allows one to find the current at small
bias voltages and from there the linear conductance
$G=dI/dV|_{V=0}$. When the dot is symmetrically coupled to the two
leads, the unitary limit $G={2e^{2}}/{h}$ is recovered at
zero temperature. The numerical results for $G$ as a function of the
dot level $\va_d$ are shown in Fig.~\ref{fig:G_vs_ed}. As can be
noticed, the method underestimates $G$ and the unitary limit is not
exactly recovered at $\va_d=-U/2$ because of the numerical accuracy of
the self-consistent treatment.

\begin{figure}[t]
\epsfig{figure=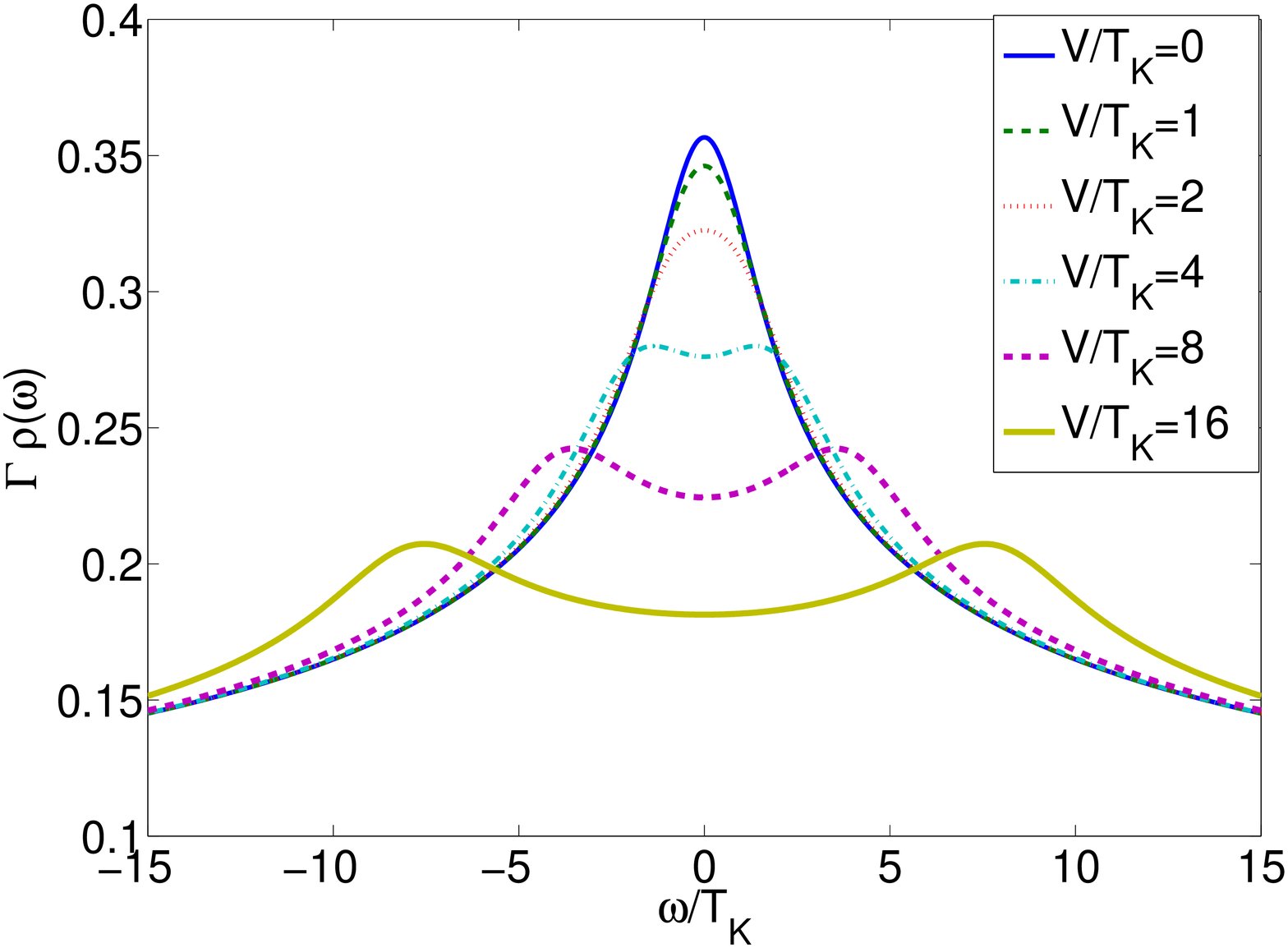,clip=,width=3.4 in}
\caption{{\protect\small {(Color online) Nonequilibrium
density of states in the particle-hole symmetric case at
$U/ \Gamma=4$ and $T/\Gamma=10^{-3}$ for different values of the
bias voltage $V$. The chemical potentials of the two leads are
taken equal to $\mu_{L/R}=\pm V/2$. The Kondo resonance peak
splits into two side peaks located at $\omega=\pm V/2$, i.e. at
the positions of the left- and right-lead chemical potentials. For
convenience we choose to represent the two energy scales (energy
$\omega$ and bias voltage $V$) as normalized by the factor
$T_{K}^{-1}$.}}}
 \label{fig:DOSvsV}
\end{figure}

\subsection{Out of equilibrium\label{sec:neq}}

\subsubsection{Differential conductance}

\begin{figure}[t]
\begin{center}
\epsfig{figure=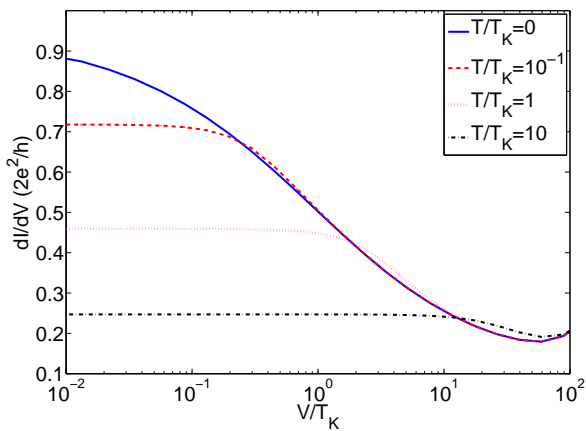,clip=,width=3.4 in}
\end{center}
\vspace{-0.3cm} \caption{\protect\small {(Color online) Differential conductance $dI/dV$ versus the bias voltage $V$ in
the particle-hole symmetric case at $U/\Gamma=4$ for different
values of temperature. The curves show a zero-bias peak, followed by the beginning of a broad Coulomb peak at large bias voltage. The differential conductance is reduced when either the temperature or the bias voltage increases,
suggesting that the Kondo effect is suppressed by temperature or
nonequilibrium effects. }} \label{fig:diff_cond_vs_T}
\end{figure}

\begin{figure}[t]
\vspace{.2cm}
\begin{center}
\epsfig{figure=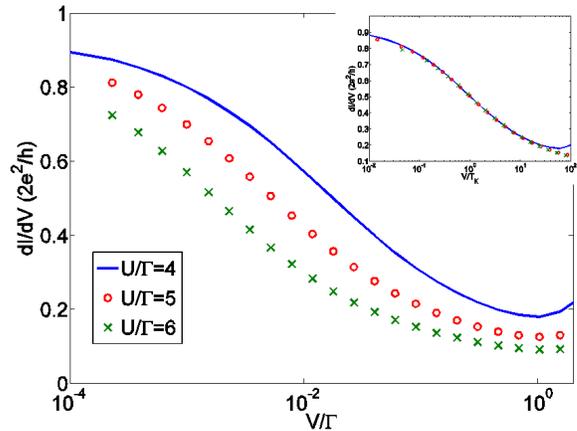,clip=,width=3.1 in}
\end{center}
\vspace{-0.4cm} 
\caption{\protect\small {(Color online) Differential conductance $dI/dV$ versus the bias voltage $V$ at $T/T_K=0.1$ in
the particle-hole symmetric case for different
values of $U$. The inset shows that the differential conductance as a function of normalized bias voltage $V/T_K$ scales to a single universal curve $dI/dV = f(V/T_K)$. At higher voltages, the
universal behavior is destroyed by a broad peak resulting from
charge fluctuations.}} \label{fig:diff_cond_vs_U}
\end{figure}

Out of equilibrium, the density of states in the dot is greatly
influenced by the bias voltage or the difference between the chemical potentials of the leads. Fig.~\ref{fig:DOSvsV} reports our results
for the nonequilibrium density of states, again in the
particle-hole symmetric case at $T/\Gamma=10^{-3}$ and $U/
\Gamma=4$ for different values of the bias voltage $V$. In
contrast with the equilibrium situation, the Kondo resonance peak splits into two
lower peaks pinned at the chemical
potentials of the two leads. The reason is that the transitions between the ground
state and the excited states of the dot are now mediated by the
conduction electrons with energies lying close to the left- and
right-lead chemical potentials.\

We then compute the differential conductance as a function of bias
voltage for different temperatures and plot the results in
Fig.~\ref{fig:diff_cond_vs_T}. At low temperatures, the bias voltage
dependence of the differential conductance shows a narrow peak at low bias (zero-bias anomaly) reflecting the Kondo effect (mind the logarithmic horizontal axis), followed by a Coulomb peak centered around the value of the dot level energy. Increasing temperature diminishes the intensity of the zero-bias peak, meaning that the Kondo effect is destroyed by temperature.

In order to discuss the universality of the dependence of the differential conductance on the bias voltage, we plot in
Fig.~\ref{fig:diff_cond_vs_U} the results obtained at zero temperature for different
values of the Coulomb interaction $U$. In the inset, the differential conductance is found to be a universal function of the renormalized bias voltage $V/T_K$, independent of other energy scales such as $U$ or $\Gamma$. This one-parameter scaling is obtained over a large range of $V$. Universality is lost around $V > 10 T_K$. Note that when $V/T_K< 0.1$, the unitary limit is not completely recovered
for the differential conductance due to the numerical accuracy in the
self-consistency treatment, as was already mentioned before.

The physical origin of the destruction of the Kondo effect is the decoherence rates induced by the voltage-driven current. As
we discussed in Sec.\ref{sec:4order_lifetimes}, these effects are well described by our
EOM approach since it incorporates higher-order
terms in $t_\sigma$. They originate physically from the
energy-conserving processes in which one electron hops onto the dot
from the higher chemical potential while another electron hops
out to the lower chemical potential. Since the
processes involve two electrons hopping in and out, the
lowest-order contribution is fourth order in $t_\sigma$. These
rates broaden and diminish the Kondo resonance peaks in the
density of states as the bias voltage increases, see
Fig.~\ref{fig:DOSvsV}. Their effect leads to a decrease in the
differential conductance when $V\ge T_K$, as was shown in
Figures \ref{fig:diff_cond_vs_T} and \ref{fig:diff_cond_vs_U}. We will analyse this in more detail later in this section.\

\begin{figure}[t]
\begin{center}
\epsfig{figure=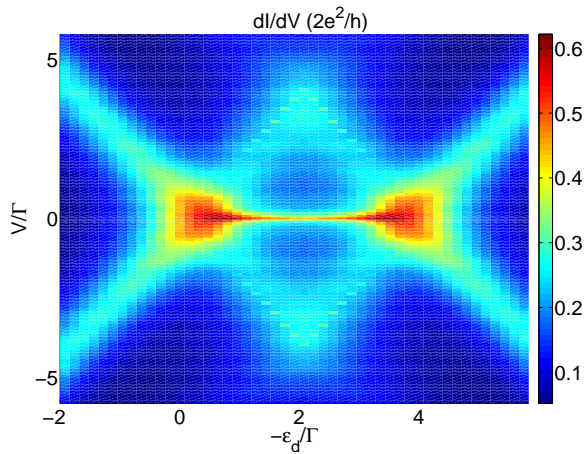,clip=,width=3.4 in}
\end{center}
\vspace{-0.3cm} 
\caption{Color plot of the differential conductance $dI/dV$ as a function of bias voltage $V$ and dot-level energy $\va_d$ for $U/\Gamma=4$ and $T/\Gamma=10^{-3}$. The contour of the Coulomb peaks delimits the Coulomb blockade diamond, separating areas with well-defined dot occupation number $\mathcal{N}$ ranging from $0,~1$ to $2$ at low $V$, and areas of charge fluctuations at high $V$. In the $\mathcal{N}=1$ central valley, $dI/dV$ shows a zero-bias peak typical of the
Kondo effect. } 
\label{fig:diff_cond_3D_Uis4}
\end{figure}

To further demonstrate that the method can work in a
wide range of parameters, we report in
Fig.\ref{fig:diff_cond_3D_Uis4} the differential conductance
in the $V-\va_d$ plane in a 3D-plot. The figure shows the usual
Coulomb diamond defining inside the Coulomb blockade regime
$\mathcal{N}=1$, where $\mathcal{N}$ is the total occupation number
in the dot ($\mathcal{N}=\sum_{\sigma}n_{\sigma}$). The
boundaries of the Coulomb diamond are related to the
values of the renormalized dot level energies $\pm\varepsilon_{d}$ and
$\pm\varepsilon_{d}+U$ (with some additional renormalization effects
in the mixed valence regime). Within the Coulomb diamond along the
$V=0$ line, one can clearly see the zero-bias peak as discussed in
Fig.~\ref{fig:G_vs_ed}. At zero temperature, the unitary limit $2e^2/h$ is almost reached at the particle-hole symmetric point
($\va_d=-U/2$). When the
temperature is increased, the zero-bias differential conductance
decreases at this point, leaving aside two broad Coulomb peaks corresponding to the alignment of the dot level energy with
the chemical potentials in the leads ($\va_d=-U$ and
$\va_d=0$).

\subsubsection{Comparison with other studies}
\begin{figure}[t]
\begin{center}
\epsfig{figure=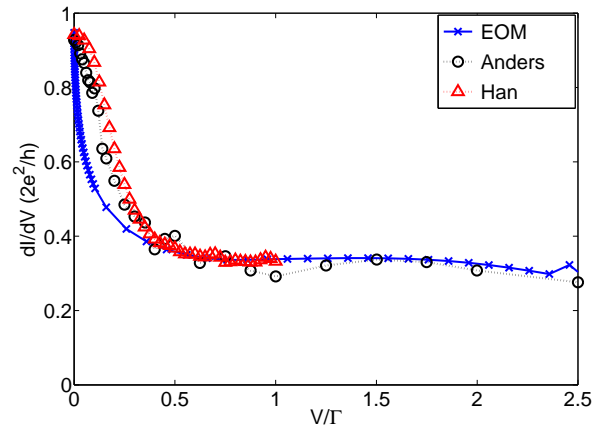,clip=,width=3.4 in}
\end{center}
\vspace{-0.7cm}
\caption{Comparison of the differential conductance $dI/dV$ as a
function of bias voltage $V$ with the results obtained by
Anders\cite{Anders2008} and Han\cite{Han2009} for
$-2\va_d/\Gamma=U/\Gamma=2.5$ and $T/\Gamma=0.008\ll T_K/\Gamma$  (We are grateful to J.E. Han for providing us with his data points). Our curve is plotted for $T=0$ in order to compare the three results in the strong coupling regime. $dI/dV$ at small $V$ is slightly different in the EOM approach because
its value for $T_K$ is smaller. At high bias voltage, the results of the three approaches agree perfectly. A little unphysical bump is observed for the EOM result at $V=U=2.5\Gamma$, when the chemical potentials of the leads are aligned with the resonant levels of the dot ($\mu_L=\va_d+U$, $\mu_R=\va_d$).}
\label{fig:compare_AndersHan}
\end{figure}

We compare our results for the differential conductance with those
obtained by other groups using time-dependent Numerical Renormalisation Group\cite{Anders2008} and an imaginary-time theory solved by using Quantum Monte Carlo\cite{HanHeary2007,Han2009}, and plot the results
obtained for the bias voltage dependence of the differential
conductance at zero temperature for comparison (Fig.~\ref{fig:compare_AndersHan}).
One finds a qualitative agreement at low bias voltages, when the
system is in the strong coupling regime. In that regime, our method slightly
underestimates $dI/dV$ because it gives a smaller Kondo scale. The three curves join at higher bias voltages, where a
quantitative agreement is found. A little local bump is observed
for the EOM result at $V=U$, when the chemical potentials of the
leads are aligned with the resonant levels of the dot
($\mu_L=\va_d+U$, $\mu_R=\va_d$). This is related to the fact that
we used the bare $D_\sa(\va)$ functions (\ref{eq:Dfunc}) in the
non-Kondo regime, leading to divergence at
$\mu_{L(R)}=\left\{\va_d,\va_d+U\right\}$. This bump can be
smeared out by introducing a finite width of order $\Gamma$ into
the $D_\sa(\va)$ functions, as can be physically originated from charge
fluctuations on the dot resonant levels.

\subsubsection{Crossover from strong coupling to weak coupling regime \label{num_decoh}}

When a bias voltage is applied to the leads, it is interesting to know whether or not the decoherence effects induced by the voltage-driven current (cf.
Sec.\ref{sec:4order_lifetimes}) may drive the system from strong to weak coupling regime. We
point out that this problem has been discussed in previous studies
for the Kondo model using either a perturbative renormalization
approach\cite{Coleman2001} or a slave-boson technique within non-crossing
approximation\cite{Rosch2001}. We would like to tackle this question for the
Anderson model with two leads using the EOM scheme.\

\begin{figure}[t]
 \centering \subfigure[] {\ \label{fig:figure_SCtoWC}
\includegraphics[width=0.47\textwidth]{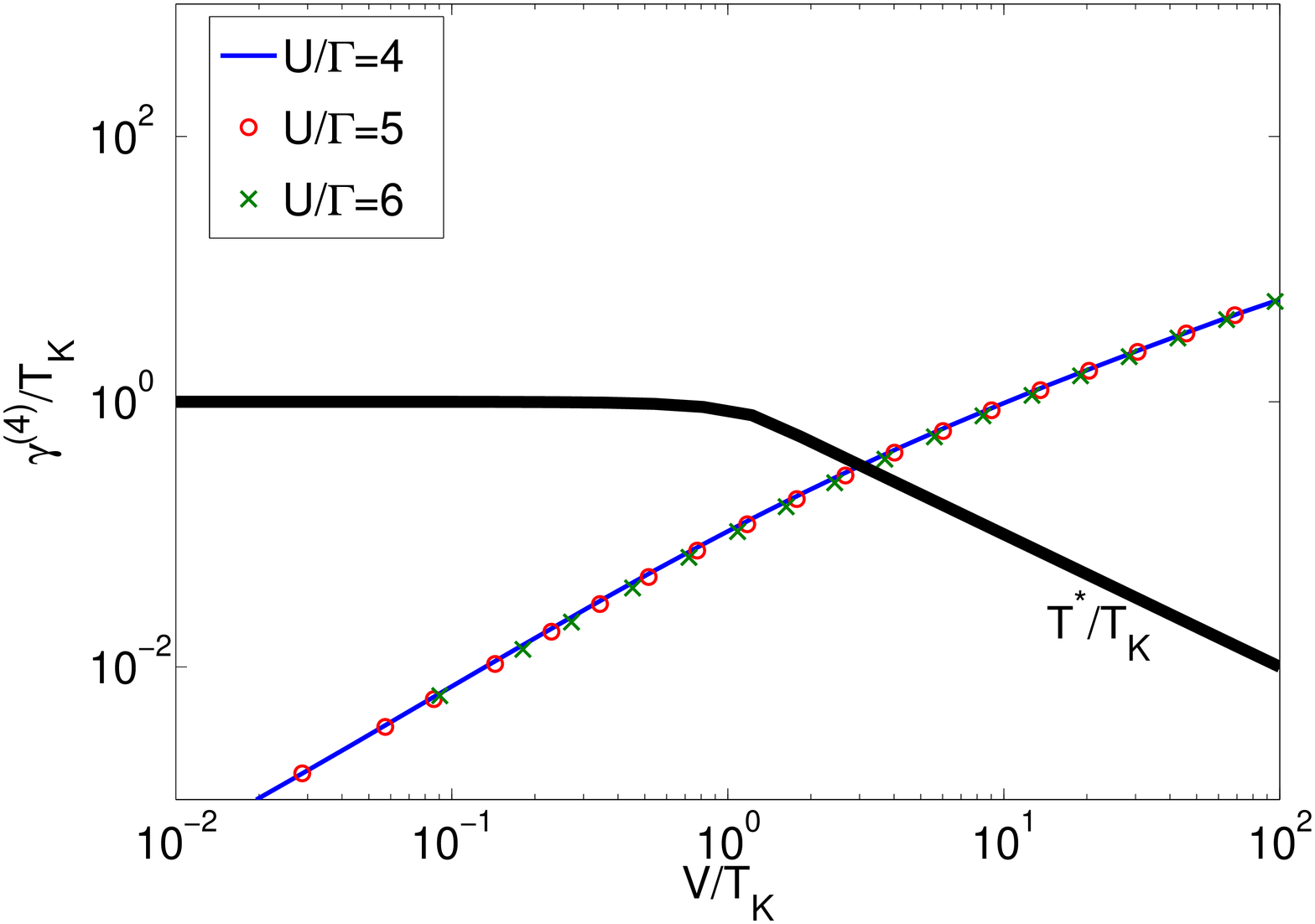}}\\
\subfigure[] {\ \label{fig:SC_to_WC}
\includegraphics[width=0.45\textwidth]{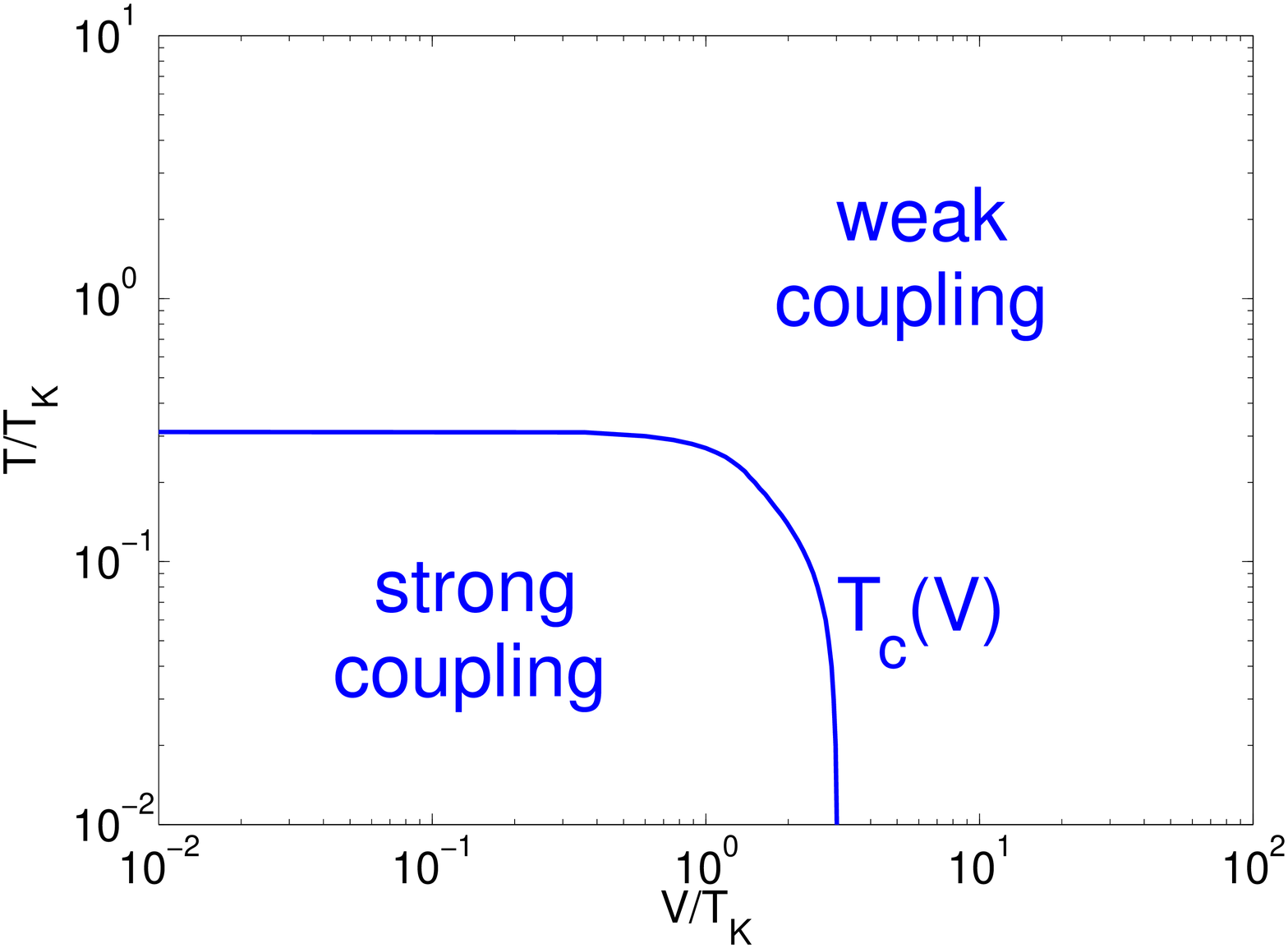} }\\
\caption{\protect\small {(Color online) 
(a) Decoherence rate $\gamma^{(4)}$ and characteristic energy scale $T^*$ versus the normalized bias voltage $V/T_K$ at $T/T_K=10^{-1}$ in the particle-hole symmetric case for several values of $U/\Gamma$. $\gamma^{(4)}/T_K$ is a universal function of $V/T_K$ over a large range of $V$. The comparison of both energy scales ($\gamma^{(4)}$ and $T^*$) allows one to determine whether the system is in the strong coupling regime ($\gamma^{(4)}<T^*$) or weak coupling regime ($\gamma^{(4)}>T^*$).
(b) Stability phase diagram of the strong coupling and weak coupling regimes in the $V-T$ plane. The crossover tempeature $T_c(V)/T_K$ is a universal function of $V/T_K$.
 }}\label{kondo124}
\end{figure}

At zero temperature and at $V\gg T_K$, $\mathcal{G}^{r}_\sa(\wa)$ for the Kondo regime behaves as 
\bea
\mathcal{G}^{r}_\sa(\wa)\propto \left[{\rm ln}\left(\frac{(\wa-V/2+i\gamma^{(4)})(\wa+V/2+i\gamma^{(4)})}{T^2_K}\right) \right]^{-1} ,\label{eq:asym_G} 
\eea
where $\gamma^{(4)}$ is the decoherence rate induced by the bias voltage as given by Eq.~(\ref{eq:tau2}). $\mathcal{G}^{r}_\sa(\wa)$ given by Eq.~(\ref{eq:asym_G}) develops a pole\cite{Rosch2001} as soon as $\gamma^{(4)}$ is smaller than a characteristic energy scale $T^*$ 
\bea
T^*=\left\{ \begin{array}{r@{\quad:\quad}l}
\sqrt{T_K^2-V^2/4}& V<\sqrt{2}\, T_K\\
T_K^2/V& V>\sqrt{2} \, T_K \end{array}\right. \label{eq:Tstar}
\eea
From there, we define a criterion controlling the
crossover between strong coupling ($\gamma^{(4)}<T^*$) and weak coupling ($\gamma^{(4)}>T^*$) regime, as proposed in Ref.\onlinecite{Rosch2001}. In order to obtain the nontrivial decoherence rate $\gamma^{(4)}$ as a function of bias voltage $V$, we replace $D_\sa(\va)$ of the bare $ J_{\alpha\sa , \beta\sa^\prime }$ in the decoherence rate by the `dressed' $\widetilde{D}_\sa(\va)$, which is identified with the denominator of the Green function (\ref{eq:GF_01})\footnote{However, the RG analysis indicates that the flow of $J_{LL(RR)}$ is different from that of $J_{RL}$ for $\Lambda < V$, where $\Lambda$ is the cutoff, see Refs.~\onlinecite{Coleman2001,Rosch2001}. We suspect that this substitution does not work in the low-energy regime where $\wa<V$.}. 
Thus we can define a renormalized $\widetilde{J}_{\alpha\sa , \beta\sa^\prime }\equiv 2 t_{\alpha\sa } t_{\beta\sa^\prime } \widetilde{D}_\sa(\mu_\alpha)$. 
In the limit $V\gg T_K$, we find $\widetilde{J}_{\alpha\sa , \beta\sa^\prime }\propto 1/[2~{\rm ln}(V/T_K)]$ and $\gamma^{(4)}\propto V/[2~{\rm ln}(V/T_K)]^2$, which is always larger than $T^*$. 
The results for the renormalized decoherence rate $\gamma^{(4)}/T_K$ as a function of the bias voltage are reported in Fig.~\ref{fig:figure_SCtoWC} for different values of $U$. Strikingly, the curves for the different values of $U$ coincide, underlining the universality of the evolution of $\gamma^{(4)}/T_K$ as a function of $V/T_K$. Combining the results for $T^*/T_K$ and $\gamma^{(4)}/T_K$, one can derive the universal crossover bias voltage $V_c/T_K$ from strong to weak coupling regime.

At finite temperatures, the derivation for $T^*$ is the same except for replacing $\gamma^{(4)}\rightarrow \sqrt{(\gamma^{(4)})^2+\pi^2T^2}$ in Eq.~(\ref{eq:asym_G}). The results are plotted in Fig.~\ref{fig:SC_to_WC} in the $V-T$ plane, displaying the crossover from strong coupling to weak
coupling regime. Although the physical mechanism at the origin of the crossover is different, both bias voltage and temperature
drive the system to the weak coupling regime.

\subsubsection{Nonequilibrium occupation number in the dot}
Typically, at equilibrium and for zero temperature, $\lan n_\sa\ran $ is mainly determined by the weight of the broad resonance peak far below the Fermi level. The narrow Kondo resonance near the Fermi energy has little weight in comparison. Thus, even if a EOM approach in a certain approximation scheme happens to describe only qualitatively Kondo physics, it is able to determine numerically the occupation number that agrees reasonably well with the Bethe ansatz or NRG.\

When the system is driven out of equilibrium, the problem becomes more complicated as one should use lesser Green functions instead of retarded ones to compute the expectation values. As discussed in Sec.\ref{sec:thermal average}, the only place where this cannot be circumvented is precisely for the dot occupation number $\lan n_\sa\ran $ appearing in the Green function (\ref{eq:GF_01}). A rigorous treatment would require to compute the lesser Green function $\mathcal{G}^<_\sa(\wa)$ and then obtain $\lan n_\sa\ran $ according to Eq.~(\ref{eq:def_ocuN}), which is beyond the scope of this work. As described in Sec.\ref{sec:thermal average}, we used instead the Ng ansatz to compute the dot occupation number. On the other hand, if we consider the particle-hole symmetric case ($\va_d=-U/2$), with a symmetric bias voltage setting $[\mu_L,\mu_R] = [V/2,-V/2]$, one obtains $\lan n_\sa\ran=1/2$ by symmetry. However, we noticed that the calculation of the occupation number by applying the Ng ansatz to our Green function leads to slight deviation from $\lan n_\sa\ran=1/2$. In \ref{App:symmH}, we show how to solve this problem.\

\begin{figure}[t]
\begin{center}
\epsfig{figure=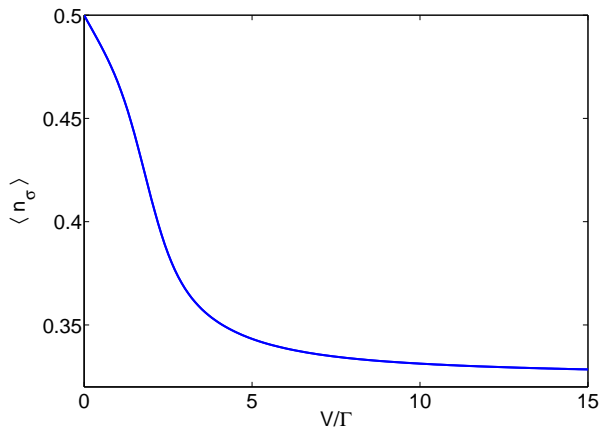,clip=,width=3.4 in}
\end{center}
\vspace{-0.6cm}
    \caption{{\protect\small {(Color online) Occupation number in the dot $\langle n_\sa \rangle$ versus the bias voltage $V$ in the particle-hole symmetric case for $U/\Gamma$ and $T/\Gamma=8.5 10^{-4}$, and under an asymmetric bias voltage setting $\mu_L =0$ and $\mu_R=-V$.}}}
    \label{fig:occnum_vs_bias}
\end{figure}

On the other hand, for an asymmetric bias voltage setting, the occupation number $\lan n_\sa\ran $ is no longer fixed by symmetry arguments. Let us take $[\mu_L,\mu_R] = [0,-V]$, the bias voltage dependence of the occupation number is shown in Fig.~\ref{fig:occnum_vs_bias}. As $V$ increases, $\lan n_\sa\ran $ decreases rapidly till $V$ passes the dot-level energy $U/2$ and comes to stabilize at large $V$. This can be qualitatively explained by the fact that at large $V$, the current through the dot no
longer increases monotonously with the bias voltage and
reaches a horizontal asymptote. This makes the occupation number insensitive to the bias voltage.\

\section{Conclusions\label{secConclusions}}

We have presented a study of the nonequilibrium
effects in the two-lead Anderson model. The calculations have been
performed within a self-consistent EOM approach generalized to the
nonequilibrium situation. The approximation scheme presented in this paper goes
beyond the previous truncations of the equations of motion done at the
second or fourth order in tunneling $t_\sigma$, by including contributions from the next orders (sixth order), which have
been shown to be of great importance out of equilibrium.\

The situation at equilibrium is used as a benchmark for the
approximation. The results for the density of states and the linear
conductance at equilibrium are found to be quantitatively improved
compared to those obtained by the EOM method using the Lacroix approximation. In the Kondo regime for instance, the
Kondo temperature $T_K$ is closer to the exact results found with the Bethe ansatz and NRG,  and non longer vanishes in the
particle-hole symmetric case. When the dot is symmetrically coupled to the leads, the linear conductance reaches
its unitary limit $2e^2/h$ at zero temperature in the Kondo regime.\

We have also computed the nonequilibrium decoherence rate $\gamma^{(4)}$ in the Kondo regime. At $T\gg T_K$, $\gamma^{(4)}/T_K$ is found to be a universal increasing
function of the normalized bias voltage $V/T_K$, depending on a
single energy scale $T_K$. The scaling law holds over a wide range
of $V$ going from $0$ to $100T_K$. At low temperature, the
density of states shows a splitting of the Kondo resonance into two
peaks, pinned at the chemical potentials of the two leads. The
height of the two peaks is controlled by the decoherence rate.\

As far as the differential conductance is concerned, it shows a
zero-bias peak at low temperature, followed
by a broad Coulomb peak at larger bias voltage. At low bias
voltage, the differential conductance also obeys a universal scaling law as a function of $V/T_K$.
Finally we have discussed the role played by the decoherence rate
$\gamma^{(4)}$ in driving the system from the strong coupling to the
weak coupling regime. We have derived the crossover line $T_c(V)$ separating the strong coupling regime to the
weak coupling regime.

\begin{acknowledgments}
We would like to thank A. Cr\'{e}pieux, P. Durganandini, W.F.
Tsai, and L.I. Glazman for valuable discussion and comments. We are also grateful
to J.E. Han for providing us with his data points. Work has been
supported by the contract ANR-05-Nano-050-S2 "QuSpins".\

$^*$ Also at the Centre National de la Recherche Scientifique (CNRS), France.
\end{acknowledgments}


\newpage
\renewcommand{\thesection}{\mbox{Appendix~\Roman{section}}} 
\setcounter{section}{0}

\renewcommand{\theequation}{\mbox{A.\arabic{equation}}} 
\setcounter{equation}{0} 

\section{Derivation of equations of motion for finite Coulomb interaction\label{App:EOM}}
\subsection{\label{App:EOM_l}}

In Section \ref{secEOM}, we have derived the first equations of motion. In this appendix, we present the detailed derivation of the higher hierarchy of equations and the decoupling scheme that follows. We derive the EOM of the higher Green functions on the right-hand side of Eq.~(\ref{eq:f4}) by using Eq.~(\ref{eq:Zubarov}). They are
\begin{widetext}
\begin{subeqnarray}
\label{app:eq:f4}
\wa_{:k}\langle\langle n_\bsa c_{k\sa}\rangle\rangle &=&t_{\sa} \langle\langle n_\bsa f_{\sa}\rangle\rangle +\sum_{k^\prime}t_\bsa \Big[\langle\langle
f_\bsa ^{\dag}c_{ k^\prime \bsa}c_{k\sa}\rangle\rangle  -\langle\langle
c_{ k^\prime \bsa}^{\dag}f_\bsa c_{k\sa}\rangle\rangle  \Big],\slabel{eq:f4a} \\
\wa_{\bsa:\sa k} \langle\langle f_\bsa^{\dag}c_{k\bsa}f_{\sa}\rangle\rangle &=&
\langle f_\bsa^{\dag} c_{k\bsa}\rangle  + t_{\bsa} \langle\langle n_\bsa f_{\sa}\rangle\rangle +\sum_{ k^\prime}\Big[ t_{\sa}\langle\langle
f_\bsa^{\dag}c_{k\bsa}c_{ k^\prime \sa}\rangle\rangle -t_{\bsa}\langle\langle
c_{ k^\prime \bsa}^{\dag}c_{k\bsa}f_{\sa}\rangle\rangle \Big],\slabel{eq:f4b}\\
\left(\wa_{k:\sa\bsa} - U\right) \langle\langle c_{k\bsa}^{\dag}f_\bsa f_{\sa}\rangle\rangle &=&
\langle c_{k\bsa}^{\dag} f_\bsa\rangle  -t_\bsa\langle\langle
n_\bsa f_{\sa}\rangle\rangle +\sum_{ k^\prime} \Big[t_\bsa \langle\langle c_{k\bsa}^{\dag}c_{ k^\prime \bsa}f_{\sa}\rangle\rangle  +t_{\sa}\langle\langle c_{k\bsa}^{\dag}f_\bsa c_{ k^\prime \sa}\rangle\rangle
\Big] , \slabel{eq:f4c}
\end{subeqnarray}
\end{widetext}
where we denote for a shorthand
\bea
\wa_{\alpha \beta \cdots : a b \cdots} &\equiv& \wa+\va_\alpha+\va_\beta+\cdots-\va_a-\va_b-\cdots  , \nn
\eea
with $\left\{\alpha \beta \cdots , a b \cdots\right\}$ being any set of parameters within $k$'s and $\sa$'s. \

For most practical purposes, the truncation is performed at this level by decoupling the second-order terms on the right-hand side of Eqs.~(\ref{app:eq:f4}), see the paragraph in the main text after Eqs.~(\ref{eq:f42}). This is done by grouping all possible same-spin pairs of lead ($c$) and dot ($f$) electron operators since we assume the spin quantum number is preserved through tunneling: any correlation between electrons of different spins has to come via the Coulomb interaction. A solution obtained at this level by neglecting the connected Green functions is exact to second order in hybridization\cite{Kashcheyevs2006}. Numerous such solutions can be found in the literature\cite{Lacroix1981,MWL1991,Appelbaum1969,Wohlman2005}, with some more elaborate than the others.

However, as discussed in the main text, stopping the flow at this point will raise terms suffering from logarithmic divergences. In the following, we show how to go beyond the second-order to derive higher equations of motion exact up to the fourth order. To begin with, we consider the following second-generation EOM:
\begin{widetext}
\begin{subeqnarray}
\label{app:eq:e02}
\wa_{\bsa:k\kp}\langle\langle
 f_\bsa ^{\dag}c_{ k^\prime \bsa} c_{k\sa}\rangle\rangle &=& -U\llan n_\sa  f_\bsa ^{\dag} c_{ k^\prime \bsa}  c_{k\sa} \rran + t_\bsa \llan n_\bsa  c_{k\sa}\rran + t_\sa \llan  f_\bsa ^{\dag}c_{ k^\prime \bsa}  f_{\sa}\rran - \sum_{\kpp }t_\bsa \llan  c_{\kpp\bsa} ^{\dag}c_{ k^\prime \bsa}  c_{k\sa}\rran,  \slabel{eq:e02}\\
\wa_{\kp:\sa k}\langle\langle
 c_{ k^\prime \bsa}^{\dag}c_{k\bsa}  f_{\sa}\rangle\rangle &=&f_{\kp k}^\bsa +U \langle\langle
n_\bsa c_{ k^\prime \bsa}^{\dag}c_{k\bsa}  f_{\sa}\rangle\rangle  +t_\bsa \langle\langle
 c_{ k^\prime \bsa}^{\dag}f_{\bsa}  f_{\sa}\rangle\rangle - t_\bsa\langle\langle
 f_{ \bsa}^{\dag}c_{k\bsa}  f_{\sa}\rangle\rangle +   \sum_\kpp t_\sa\langle\langle
 c_{ k^\prime \bsa}^{\dag}c_{k\bsa}  c_{\kpp \sa}\rangle\rangle,  \slabel{eq:e03}\\
\wa_{\kp:\bsa k}\langle\langle
 c_{ k^\prime \bsa}^{\dag}f_{\bsa}  c_{k\sa}\rangle\rangle &=&U\langle\langle n_\sa
  c_{ k^\prime \bsa}^{\dag}f_{\bsa}  c_{k\sa}\rangle\rangle -t_\bsa \llan  n_\bsa  c_{k\sa},f_{\sa}^{\dag}\rran+ t_\sa \langle\langle
 c_{ k^\prime \bsa}^{\dag}f_{\bsa}  f_{\sa}\rangle\rangle +  \sum_\kpp t_\bsa\langle\langle
 c_{ k^\prime \bsa}^{\dag}c_{\kpp\bsa}  c_{k\sa}\rangle\rangle. \slabel{eq:e04}
\end{subeqnarray}
\end{widetext}
where we denote $f_{\kp k}^\sa  \equiv  \lan c_{\kp\sa}^\dag c_{k\sa}\ran.$

We proceed to insert the above Eqs.~(\ref{app:eq:e02}) into Eqs.~(\ref{app:eq:f4}). The right-hand side of the latter equations involve new Green functions generated via the Coulomb interaction and others of the same hierarchy as the left-hand side. The latter Green functions can either move to the left-hand side or vanish in the wide-band limit since upon summing over $k$, all denominators have poles in the upper half complex plane. Furthermore, we decouple $\llan c^\dag_\bsa c_\bsa c_\sa\rran \approx  \lan c^\dag_\bsa c_\bsa\ran \llan c_\sa\rran$ and then use Eq.~(\ref{eq:f2}), since this decoupling should be exact up to order of $t_\sa^4$ and for another reason which will be clear later.
We end up with
\begin{widetext}
\begin{subeqnarray}
\label{app:eq:f402}
\left[\wa_{:k}-\Sigma_{1\sa}(\wa_{:k})\right]\langle\langle n_\bsa c_{k\sa}\rangle\rangle &=&t_{\sa} \langle\langle n_\bsa f_{\sa}\rangle\rangle -U\sum_\kp t_\bsa \left[\wa_{\bsa:k\kp}^{-1}\llan n_\sa  f_\bsa ^{\dag}c_{ k^\prime \bsa}  c_{k\sa} \rran  +\wa_{\kp:\bsa k}^{-1}\langle\langle n_\sa
 c_{ k^\prime \bsa}^{\dag}f_{\bsa}  c_{k\sa}\rangle\rangle \right],\slabel{eq:f4a02}\\
\left[\wa_{\bsa:\sa k}-\Sigma_{2\sa}(\wa_{:k})\right]\langle\langle f_\bsa^{\dag}c_{k\bsa}f_{\sa}\rangle\rangle &=&
\langle f_\bsa^{\dag} c_{k\bsa}\rangle  + t_{\bsa} \langle\langle n_\bsa f_{\sa}\rangle\rangle  -\sum_\kp  t_\bsa \wa_{\kp:\sa k}^{-1} f_{\kp k}^\bsa \left[1 + \Sigma^0_\sa(\wa)\llan f_\sa\rran  \right]    \nonumber\\ &&
-U\sum_\kp  \left[t_\sa\wa_{\bsa:k\kp}^{-1}\llan n_\sa  f_\bsa ^{\dag}c_{ k \bsa}  c_{\kp\sa} \rran  +t_\bsa\wa_{\kp:\sa k}^{-1}\langle\langle n_\bsa
 c_{ k^\prime \bsa}^{\dag}c_{k\bsa}  f_{\sa}\rangle\rangle \right],\slabel{eq:f4b02}\\
\left[\wa_{k:\sa\bsa}-U-\Sigma_{3\sa}(\wa_{k:})\right]\langle\langle c_{k\bsa}^{\dag}f_\bsa f_{\sa}\rangle\rangle &=&
\langle c_{k\bsa}^{\dag} f_\bsa\rangle  -t_\bsa\langle\langle
n_\bsa f_{\sa}\rangle\rangle + \sum_\kp t_\bsa \wa_{k:\sa\kp}^{-1}  f_{k\kp }^\bsa \left[1 + \Sigma^0_\sa(\wa)\llan f_\sa\rran  \right]
\nn\\
&&+U\sum_\kp  \left[t_\sa\wa_{k:\bsa\kp}^{-1}\llan n_\sa  c_{ k \bsa}^{\dag}f_\bsa c_{\kp\sa}\rran  +t_\bsa\wa_{k:\sa\kp}^{-1}\langle\langle n_\bsa
 c_{ k \bsa}^{\dag}c_{\kp\bsa}  f_{\sa}\rangle\rangle \right].\slabel{eq:f4c02}
\end{subeqnarray}
\end{widetext}
where
\bea
\Sigma^0_\sa(\wa)&=& \sum_k t_\sa^2 \wa_{:k}^{-1},\\
\Sigma_{1\sa}(\wa_{:k})&=&\sum_{\kp}t_\bsa^2\left[\wa_{\bsa:k\kp }^{-1}+\wa_{\kp:\bsa k}^{-1}\right],\label{eq:defunction_001}\\
\Sigma_{2\sa}(\wa_{:k})&=&\sum_{\kp}\left[t_\sa^2\wa_{\bsa: k\kp}^{-1}+t_\bsa^2\wa_{\kp:\sa k}^{-1}\right],\label{eq:defunction_002}\\
\Sigma_{3\sa}(\wa_{k:})&=&\sum_{\kp}\left[t_\bsa^2 \wa_{k:\sa \kp}^{-1}+t_\sa^2\wa_{k:\bsa\kp }^{-1}\right].\label{eq:defunction_003}
\eea

It is interesting to notice that at this level the prefactor of the Green functions on the left-hand side acquires non-interacting self-energy terms, while new Green functions remain on the right-hand side. This allows us to focus on the new Green functions, which is of most importance. \
We list below these third-generation equations of motion,
\begin{widetext}
\begin{subeqnarray}
\label{app:eq:aee04}
\left(\wa_{\bsa:\kp k} + U \right) \llan n_\sa f_\bsa ^{\dag}c_{ k \bsa}c_{\kp\sa} \rran &=& -\lan f_\bsa ^{\dag}c_{ k  \bsa}f_\sa^{\dag}c_{\kp\sa} \ran + t_\bsa \llan n_\sa n_\bsa c_{\kp\sa} \rran- \sum_\kpp t_\sa \llan c_{\kpp \sa}^\dag f_\sa f_\bsa ^{\dag}c_{ k  \bsa}c_{\kp\sa} \rran \nn\\
&&+\sum_\kpp t_\sa \llan f_\sa^\dag c_{\kpp \sa} f_\bsa ^{\dag}c_{ k \bsa}c_{\kp\sa} \rran -\sum_\kpp t_\bsa \llan n_\sa c_{\kpp \bsa}^\dag c_{ k \bsa}c_{\kp\sa}\rran, \slabel{eq:aee04}
\\
\left(\wa_{\kp:\bsa k}-U\right)\langle\langle n_\sa
c_{ k^\prime \bsa}^{\dag}f_{\bsa}c_{k\sa}\rangle\rangle &=& -\lan c_{ k^\prime \bsa}^{\dag}f_{\bsa}f_{\sa}^{\dag} c_{k\sa} \ran -t_\bsa \llan n_\sa n_\bsa c_{k \sa}  \rran-\sum_\kpp t_\sa \llan c_{\kpp \sa}^\dag f_\sa c_{ k^\prime \bsa}^{\dag}f_{\bsa}c_{k\sa} \rran \nn\\
&& + \sum_\kpp t_\sa \llan f_\sa^\dag c_{\kpp \sa}  c_{ k^\prime \bsa}^{\dag}f_{\bsa}c_{k\sa} \rran  +  \sum_\kpp t_\bsa \llan n_\sa
c_{ k^\prime \bsa}^{\dag}c_{\kpp \bsa}c_{k\sa}\rran,\slabel{eq:aee05}\\
\left(\wa_{\kp:\sa k} - U\right)\langle\langle
n_\bsa c_{ k^\prime \bsa}^{\dag}c_{k\bsa}f_{\sa}\rangle\rangle &=& \lan n_\bsa c_{ k^\prime \bsa}^{\dag}c_{k\bsa}\ran -\sum_\kpp t_\bsa\llan c_{\kpp \bsa}^\dag f_\bsa c_{ k^\prime \bsa}^{\dag}c_{k\bsa}f_{\sa}\rran \nn\\
&&-\sum_\kpp t_\bsa\llan  f_\bsa^\dag c_{ k^\prime \bsa}^{\dag}c_{\kpp \bsa}c_{k\bsa}f_{\sa}\rran + \sum_\kpp t_\sa\llan n_\bsa c_{ k^\prime \bsa}^{\dag}c_{k\bsa}c_{\kpp\sa}\rran. \slabel{eq:aee06}
\end{subeqnarray}
Up to order $t_\sa^4$, we decouple Eqs.~(\ref{app:eq:aee04}) by considering the following decouplings:
\begin{subeqnarray}
\lan f_\bsa ^{\dag}c_{ k  \bsa}f_\sa^{\dag}c_{\kp\sa} \ran &\approx&\lan f_\bsa ^{\dag}c_{ k  \bsa}\ran \lan f_\sa^{\dag}c_{\kp\sa} \ran,\\
\lan n_\bsa c_{ k^\prime \bsa}^{\dag}c_{k\bsa}\ran &\approx&\lan n_\bsa\ran f_{\kp k}^\bsa -\lan f_\bsa^\dag c_{k\bsa}\ran\lan c_{\kp \bsa}^\dag f_\bsa\ran,\\
\lan c_{ \kp \bsa}^{\dag}f_{\bsa}f_{\sa}^{\dag} c_{k\sa} \ran &\approx&\lan c_{ k^\prime \bsa}^{\dag}f_{\bsa}\ran \lan f_{\sa}^{\dag} c_{k\sa} \ran,\\
\llan c_{\kpp \sa}^\dag f_\sa f_\bsa ^{\dag}c_{ k  \bsa}c_{\kp\sa} \rran &\approx&-f_{\kpp \kp}^\sa\llan  f_\bsa ^{\dag}c_{ k  \bsa}  f_\sa\rran-\lan f_\bsa ^{\dag}c_{ k  \bsa}\ran \llan c_{\kpp \sa}^\dag  c_{\kp\sa} f_\sa\rran+f_{\kpp \kp}^\sa\lan f_\bsa ^{\dag}c_{ k  \bsa}\ran \llan  f_\sa  \rran,\slabel{eq:dec01}\\
\llan f_\sa^\dag c_{\kpp \sa} f_\bsa ^{\dag}c_{ k \bsa}c_{\kp\sa} \rran&\approx& \lan  f_\bsa ^{\dag}c_{ k \bsa}\ran \llan f_\sa^\dag c_{\kpp \sa}c_{\kp\sa} \rran,\\
\llan n_\sa c_{\kpp \bsa}^\dag c_{ k \bsa}c_{\kp\sa}\rran&\approx& f_{\kpp k}^\bsa\llan n_\sa c_{\kp\sa}\rran ,\\
\llan c_{\kpp \sa}^\dag f_\sa c_{ k^\prime \bsa}^{\dag}f_{\bsa}c_{k\sa} \rran&\approx&-f_{\kpp k}^\sa\llan   c_{ k^\prime \bsa}^{\dag}f_{\bsa} f_\sa\rran-\lan c_{ k^\prime \bsa}^{\dag}f_{\bsa}\ran\llan c_{\kpp \sa}^\dag  c_{k\sa} f_\sa\rran+f_{\kpp k}^\sa\lan c_{ k^\prime \bsa}^{\dag}f_{\bsa}\ran\llan  f_\sa \rran ,\slabel{eq:dec02}\\
\llan f_\sa^\dag c_{\kpp \sa}  c_{ k^\prime \bsa}^{\dag}f_{\bsa}c_{k\sa} \rran&\approx&\lan c_{ k^\prime \bsa}^{\dag}f_{\bsa} \ran\llan f_\sa^\dag c_{\kpp \sa} c_{k\sa} \rran,\\
\llan c_{\kpp \bsa}^\dag f_\bsa c_{ k^\prime \bsa}^{\dag}c_{k\bsa}f_{\sa}\rran&\approx& -f_{\kpp k}^\bsa \llan  c_{ k^\prime \bsa}^{\dag} f_\bsa f_{\sa}\rran+ f_{\kp k}^\bsa \llan c_{\kpp \bsa}^\dag f_\bsa f_{\sa}\rran,\\
\llan  f_\bsa^\dag c_{ k^\prime \bsa}^{\dag}c_{\kpp \bsa}c_{k\bsa}f_{\sa}\rran&\approx&f_{\kp \kpp}^\bsa \llan  f_\bsa^\dag c_{k\bsa} f_{\sa}\rran-f_{\kp k}^\bsa \llan  f_\bsa^\dag  c_{\kpp \bsa} f_{\sa}\rran,\\
\llan n_\bsa c_{ k^\prime \bsa}^{\dag}c_{k\bsa}c_{\kpp\sa}\rran&\approx&f_{\kp k}^\bsa \llan n_\bsa c_{\kpp\sa}\rran-\lan f_\bsa^\dag c_{k\bsa}\ran\llan c_{\kp\bsa}^\dag f_\bsa c_{\kpp\sa}\rran.
\end{subeqnarray}
Note that the last terms in Eqs.~(\ref{eq:dec01}, \ref{eq:dec02}) are added in order to avoid double counting from the first two terms.
On the other hand, since $\langle\langle n_\bsa n_\sa c_{k\sa}\rangle\rangle$ vanish in the wide-band limit after summing over $k$, they are removed from now on without further notice. Eqs.~(\ref{app:eq:aee04}) then become
\begin{subeqnarray}
\label{app:eq:ee04}
(\wa_{\bsa:\kp k}+U)\llan n_\sa f_\bsa ^{\dag}c_{ k \bsa}c_{\kp\sa}\rran &=&\sum_\kpp t_\sa f_{\kpp \kp}^\sa\llan   f_\bsa ^{\dag}c_{ k  \bsa}f_\sa \rran+\Big[\wa_{:\kp}\lan f_\bsa ^{\dag}c_{ k \bsa}\ran-\sum_\kpp t_\bsa f_{\kpp k }^\bsa\Big]\llan n_\sa c_{\kp\sa} \rran\nn\\
&& -\lan f_\bsa ^{\dag}c_{ k  \bsa}\ran \sum_\kpp t_\sa f_{\kpp \kp}^\sa \llan f_\sa\rran , \slabel{eq:ee04}\\
(\wa_{\kp:\bsa k}-U)\langle\langle n_\sa
c_{ k^\prime \bsa}^{\dag}f_{\bsa}c_{k\sa}\rangle\rangle &=&  \sum_\kpp t_\sa f_{\kpp k}^\sa\llan   c_{ k^\prime \bsa}^{\dag}f_{\bsa}f_\sa \rran + \Big[ \wa_{:k}\lan c_{ k^\prime \bsa}^{\dag}f_{\bsa}\ran+\sum_\kpp t_\bsa f_{\kp\kpp }^\bsa \Big]\llan n_\sa
 c_{k\sa}\rran\nn\\
&& - \lan c_{ k^\prime \bsa}^{\dag}f_{\bsa}\ran \sum_\kpp t_\sa f_{\kpp k}^\sa \llan f_\sa\rran   , \slabel{eq:ee05}\\
(\wa_{\kp:\sa  k}-U)\langle\langle
n_\bsa c_{ k^\prime \bsa}^{\dag}c_{k\bsa}f_{\sa}\rangle\rangle &=&  -\lan f_\bsa^\dag c_{k\bsa}\ran\Big[\lan c_{\kp \bsa}^\dag f_\bsa\ran+ \sum_\kpp t_\sa\llan c_{\kp\bsa}^\dag f_\bsa c_{\kpp\sa}\rran\Big] +\sum_\kpp t_\bsa f_{\kpp k}^\bsa \llan c_{ k^\prime \bsa}^{\dag} f_\bsa  f_{\sa}\rran \nn\\
&& -\sum_\kpp t_\bsa  f_{\kp\kpp }^\bsa \llan  f_\bsa^\dag c_{k\bsa}f_{\sa}\rran+ (\wa_{:\sa} -U) f_{\kp k}^\bsa \langle\langle n_\bsa f_{\sa}\rangle\rangle. \slabel{eq:ee06}
\end{subeqnarray}
To obtain Eqs.~(\ref{eq:ee04}, \ref{eq:ee05}) in a more compact form, we have used the following equation of motion,
\bd
\wa_{:k}\llan n_\sa c_{k\sa}\rran=- \lan f_\sa ^{\dag}c_{k\sa} \ran+\sum_{\kp}t_\sa\Big[\llan c_{\kp \sa}^\dag  c_{k\sa} f_\sa\rran+ \llan f_\sa^\dag c_{\kp \sa} c_{k\sa} \rran\Big]\label{eq:nc01}.
\ed
and taken Eq.~(\ref{eq:f4}) for Eq.~(\ref{eq:ee06}). Introducing Eqs.~(\ref{app:eq:ee04}) into Eqs.~(\ref{app:eq:f402}) and after some straightforward algebra, we eventually obtain
\begin{subeqnarray}
\label{app:eq:f403}
\big[\wa_{:k}-\Sigma_{1\sa}(\wa_{:k})\big]\langle\langle n_\bsa c_{k\sa}\rangle\rangle &=&t_{\sa} \langle\langle n_\bsa f_{\sa}\rangle\rangle -\sum_{\kp\kpp}t_\sa t_\bsa \Big[D_{\bsa:k \kp} \lan  f_\bsa^\dag c_{\kp\bsa}\ran+D_{\kp:\bsa k}\lan c^\dag_{\kp\bsa} f_\bsa\ran\Big]f_{\kpp k}^\sa \llan f_\sa \rran  \nn\\
&&+ \Sigma_{5\bsa}(\wa_{:k}) \langle\langle n_\sa c_{k\sa}\rangle\rangle+\sum_{\kp\kpp} t_\sa t_\bsa D_{\bsa:k \kp} f_{\kpp k}^\sa \llan   f_\bsa ^{\dag}c_{ k^\prime \bsa}f_\sa\rran\nn\\
&& + \sum_{\kp\kpp} t_\sa t_\bsa D_{\kp :\bsa k}f_{\kpp k}^\sa\llan   c_{ k^\prime \bsa}^{\dag}f_{\bsa}f_\sa  \rran ,  \slabel{eq:f4a03}\\
\big[\wa_{\bsa:k\sa}-\widehat{\Sigma}_{2\sa}(\wa_{:k})\big]\langle\langle f_\bsa^{\dag}c_{k\bsa}f_{\sa}\rangle\rangle &=&
t_{\bsa} \langle\langle n_\bsa f_{\sa}\rangle\rangle +\langle f_\bsa^{\dag} c_{k\bsa}\rangle \Big[1-\sum_{\kp} t_\bsa D_{ \kp:\sa k}\Big(\lan c_{\kp \bsa}^\dag f_\bsa\ran+ \sum_\kpp t_\sa\llan c_{\kp\bsa}^\dag f_\bsa c_{\kpp\sa}\rran\Big)  \Big]\nn\\
&& - \sum_\kp  \Big[  t_\bsa f_{\kp k}^\bsa + \lan f_\bsa^\dag c_{k\bsa} \ran\sum_\kpp t_\sa^2 D_{\bsa:\kp k}f_{\kpp \kp}^\sa  \Big] \llan f_\sa\rran  +\sum_{\kp\kpp} t_\bsa^2 D_{ \kp:\sa k}f_{\kpp k}^\bsa  \llan c_{\kp\bsa}^\dag f_\bsa f_\sa\rran \nonumber\\
&&-\sum_\kp t_\sa D_{\bsa:k \kp}\Big[\sum_{\kpp}  t_\bsa  f_{\kpp k}^\bsa-\wa_{:\kp}\langle f_\bsa^{\dag} c_{k\bsa}\rangle \Big] \llan n_\sa c_{\kp\sa}\rran ,\slabel{eq:f4ac03}\\
\big[\wa_{k:\sa\bsa}-U-\widehat{\Sigma}_{3\sa}(\wa_{k:})\big]\langle\langle c_{k\bsa}^{\dag}f_\bsa f_{\sa}\rangle\rangle &=&
  -t_\bsa\langle\langle
n_\bsa f_{\sa}\rangle\rangle+\langle c_{k\bsa}^{\dag} f_\bsa\rangle\Big[1+\sum_\kp t_\bsa D_{k:\sa \kp}\Big(\lan f_\bsa^\dag c_{\kp\bsa}\ran+\sum_\kpp t_\sa \llan f_\bsa^\dag c_{\kp\bsa} c_{\kpp\sa}\rran\Big) \Big]\nn\\
&& + \sum_\kp \Big[  t_\bsa f_{k\kp }^\bsa + \lan c_{k\bsa}^\dag f_\bsa \ran\sum_\kpp t_\sa^2 D_{k:\bsa\kp } f_{\kpp \kp}^\sa \Big] \llan f_\sa\rran  +\sum_{\kp\kpp} t^2_\bsa D_{k:\sa\kp }f_{k\kpp}^\bsa \llan f_\bsa^\dag c_{\kp\bsa}  f_\sa\rran \nn\\
&&-\sum_\kp t_\sa D_{k:\bsa\kp }\Big[\sum_\kpp  t_\bsa f_{k\kpp}^\bsa+\wa_{:\kp}\lan c_{k\bsa}^\dag f_\bsa\ran \Big]\langle\langle n_\sa c_{\kp\sa}\rangle\rangle  , \slabel{eq:f4c03}\\
\big[\wa_{:k}-\Sigma_{1\bsa}(\wa_{:k})\big]\langle\langle n_{\sa}
c_{ k \sa}\rangle\rangle
&=&-\lan f^\dag_\sa c_{k\sa}\ran +\sum_{\kp}t_\sa D_{\sa :k \kp}\Big[\sum_\kpp t_\sa f_{\kpp k}^\sa-\wa_{:\kp}\lan f_\sa^\dag c_{k\sa}\ran\Big]\llan n_\bsa  c_{\kp \sa}\rran \nn\\
&& +\Sigma_{5\sa}(\wa_{:k})\llan n_\bsa c_{ k \sa}\rran +\sum_\kp   t_\sa f_{\kp k}^\sa \llan f_\sa\rran . \slabel{eq:f4d03}
\end{subeqnarray}
where
\bea
D_{\alpha \beta \cdots : a b \cdots} &\equiv& -U \wa_{\alpha \beta \cdots : a b \cdots}^{-1} \left(\wa_{\alpha \beta \cdots : a b \cdots}\pm U\right)^{-1} ,
\eea
The sign in front of $U$ is the same as the sign in front of $\va_\sa$ in $\wa_{\alpha \beta \cdots : a b \cdots}$.
The self-energy corrections are
\bea
\Sigma_{1\bsa}(\wa_{:k})&\equiv&\sum_{\kp}t_\sa^2\left[\wa_{\sa:k \kp}^{-1}+\wa_{\kp:\sa k}^{-1} \right],\label{eq:defunction_004}\\
\widehat{\Sigma}_{2\sa}(\wa_{:k})&\equiv&\Sigma_{2\sa}(\wa_{:k})+\sum_{\kp\kpp} \Big[t_\sa^2D_{\bsa :k\kp}f_{\kpp\kp}^\sa-t_\bsa^2D_{\kp:\sa k}f_{\kp\kpp}^\bsa\Big],\label{eq:SE02}
\\
\widehat{\Sigma}_{3\sa}(\wa_{k:})&\equiv&\Sigma_{3\sa}(\wa_{k:})-\sum_{\kp\kpp} \Big[t_\sa^2D_{k:\bsa\kp}f_{\kpp\kp}^\sa+t_\bsa^2D_{k:\sa\kp}f_{\kpp\kp}^\bsa\Big].\label{eq:SE03}
\eea
We also define two following functions,
\bea
\Sigma_{5\sa}(\wa_{:k})&\equiv&\sum_{\kp\kpp}t_\sa^2 \Big[D_{ \kp:\sa k}f_{\kp\kpp}^\sa-D_{\sa:k \kp}f_{\kpp\kp}^\sa\Big]+\wa_{:k}\sum_{\kp} t_\sa \Big[D_{\kp:\sa k}\lan c^\dag_{\kp\sa} f_\sa\ran+D_{\sa:k \kp} \lan  f_\sa^\dag c_{\kp\sa}\ran\Big],\label{eq:SE04}.
\eea

After truncation there appears a new Green function $\langle\langle n_{\sa}
c_{ k \sa}\rangle\rangle$ on the right-hand side of Eqs.~(\ref{eq:f4a03}-\ref{eq:f4c03}), which has to be calculated separately. Its equation of motion, given by Eq.~(\ref{eq:f4d03}), is derived in \ref{App:EOMnc}. Note that in deriving Eqs.~(\ref{eq:f4ac03}-\ref{eq:f4d03}), we approximate
\bea
 \sum_\kp t_\bsa f_{k\kp }^\bsa\Big[ \wa_{k:\sa\kp}^{-1}\left(1+\Sigma^0_\sa(\wa)\llan f_\sa\rran \right) -  D_{ k:\sa\kp }  (\wa_{:\sa} -U)\langle\langle n_\bsa f_{\sa}\rangle\rangle\Big]&\approx& \sum_\kp t_\bsa f_{k\kp }^\bsa \llan f_\sa\rran.  \label{eq:acc03}
\eea
In doing so, we assume that the lead electron energies $\va_k,~\va_\kp\sim \mu_{L(R)}$ cancel each other; this assumption is seconded by the numerator $f_{k\kp }^\bsa= \delta_{k \kp}f^\alpha_F(\va_k)$ at zeroth order. Thus it is valid to use Eq.~(\ref{eq:f3}). This approximation should not affect the density of states around the Fermi level.

We emphasize that Eqs.~(\ref{app:eq:f403}) shown above are exact up to fourth order. The prefactor of these equations of motion acquires second-order corrections. Their imaginary part takes different values in the different regimes of the Anderson model, as is discussed in Sec.\ref{sec:analytic}.   \

\subsection{Derivation of the equation of motion of $\displaystyle \llan n_\sa c_{k\sa}\rran$ \label{App:EOMnc}}
Here we expand the equation of motion of $\langle\langle n_{\sa}
c_{ k \sa}\rran$. The equation is already given by Eq.~(\ref{eq:nc01}). We now derive the next higher equations
\begin{subeqnarray}
\label{eq:nc02}
\wa_{\sa:k \kp}\llan f_{\sa}^{\dag} c_{\kp \sa}c_{ k \sa}\rran &=& -\sum_{\kpp}t_\sa \llan c_{\kpp\sa}^{\dag} c_{\kp \sa}c_{ k \sa}\rran +t_\sa \llan n_{\sa}c_{ k \sa}\rran-t_\sa \llan n_{\sa}c_{ \kp \sa}\rran-U\llan n_\bsa f_{\sa}^{\dag} c_{\kp \sa}c_{ k \sa}\rran,\slabel{eq:nc021} \\
\wa_{\kp:\sa k}\llan  c_{\kp \sa}^{\dag}c_{ k \sa}f_{\sa}\rran  &=&f_{\kp k}^{\sa} +t_\sa \llan  n_{\sa}c_{ k \sa}\rran +\sum_\kpp t_\sa\llan c_{\kp\sa}^{\dag}c_{ k \sa} c_{\kpp \sa}\rran +U\llan  n_\bsa c_{\kp \sa}^{\dag}c_{ k \sa}f_{\sa}\rran.\slabel{eq:nc022}
\end{subeqnarray}
Similarly at this stage, we go on to decouple $\llan c_{\kpp\sa}^{\dag} c_{\kp \sa}c_{ k \sa}\rran\approx f_{\kpp\kp}^{\sa} \llan c_{ k \sa}\rran-f_{\kpp k}^{\sa} \llan c_{ \kp \sa}\rran$. After removing terms that will vanish in the wide-band limit upon summing over $k$, Eq.~(\ref{eq:nc01}) becomes
\bea
\big[\wa_{:k}-\Sigma_{1\bsa}(\wa_{:k})\big]\langle\langle n_{\sa}
c_{ k \sa}\rran &=&-\lan f^\dag_\sa c_{k\sa}\ran +\sum_{\kp}t_\sa \wa_{\kp:\sa k}^{-1}f_{\kp k}^\sa\left[ 1+\Sigma^0_\sa(\wa)\llan f_\sa\rran \right]  \nn\\
&&+U\sum_{\kp}t_\sa\left[\wa_{\kp:\sa k}^{-1}\llan  n_\bsa c_{\kp \sa}^{\dag}c_{ k \sa}f_{\sa}\rran-\wa_{\sa:k \kp}^{-1}\llan n_\bsa f_{\sa}^{\dag} c_{\kp \sa}c_{ k \sa}\rran  \right].\label{eq:nc011}
\eea
where $\displaystyle\Sigma_{1\bsa}(\wa_{:k})=\sum_{\kp}t_\sa^2\big[\wa_{\sa:k \kp}^{-1}+\wa_{\kp:\sa k}^{-1} \big]$.
Next we derive the equations of motion of the Green functions on the right-hand side of Eq.~(\ref{eq:nc011})
\begin{subeqnarray}
\label{eq:nc03}
(\wa_{\sa:k \kp}+U)\llan n_\bsa f_{\sa}^{\dag} c_{\kp \sa}c_{ k \sa}\rran &=& -\sum_\kpp t_\bsa\llan c_{\kpp\bsa}^\dag f_\bsa f_{\sa}^{\dag} c_{\kp \sa}c_{ k \sa}\rran +\sum_\kpp t_\bsa\llan f_\bsa^\dag c_{\kpp\bsa} f_{\sa}^{\dag} c_{\kp \sa}c_{ k \sa}\rran\nn\\
&&+t_\sa\llan n_\bsa n_{ \sa}c_{ k \sa}\rran-t_\sa\llan n_\bsa n_{ \sa}c_{ \kp \sa}\rran-\sum_\kpp t_\sa\llan n_\bsa c_{\kpp\sa}^{\dag} c_{\kp \sa}c_{ k \sa}\rran,\slabel{eq:nc031}\\
(\wa_{\kp:\sa k}-U)\llan  n_\bsa c_{\kp \sa}^{\dag}c_{ k \sa}f_{\sa}\rran &=&\lan n_\bsa c_{\kp \sa}^{\dag}c_{ k \sa}\ran -\sum_\kpp t_\bsa\llan c_{\kpp\bsa}^\dag f_\bsa c_{\kp \sa}^{\dag}c_{ k \sa}f_{\sa}\rran +t_\sa\llan  n_\bsa n_\sa c_{ k \sa}\rran\nn\\
&&+\sum_\kpp t_\bsa\llan f_\bsa^\dag c_{\kpp\bsa}c_{\kp \sa}^{\dag}c_{ k \sa}f_{\sa}\rran +\sum_\kpp t_\sa\llan  n_\bsa c_{\kp \sa}^{\dag}c_{ k \sa}c_{\kpp\sa}\rran.\slabel{eq:nc032}
\end{subeqnarray}
We now decouple Eqs.~(\ref{eq:nc03}) according to the following decouplings:
\begin{subeqnarray}
\lan n_\bsa c_{\kp \sa}^{\dag}c_{ k \sa}\ran&\approx& f_{\kp k}^\sa \lan n_\bsa\ran,\\
\llan c_{\kpp\bsa}^\dag f_\bsa f_{\sa}^{\dag} c_{\kp \sa}c_{ k \sa}\rran &\approx& \lan f_{\sa}^{\dag} c_{\kp \sa}\ran\llan c_{\kpp\bsa}^\dag f_\bsa c_{ k \sa}\rran-\lan f_{\sa}^{\dag} c_{k \sa}\ran\llan c_{\kpp\bsa}^\dag f_\bsa  c_{\kp \sa}\rran ,\\
\llan f_\bsa^\dag c_{\kpp\bsa} f_{\sa}^{\dag} c_{\kp \sa}c_{ k \sa}\rran&\approx& \lan f_{\sa}^{\dag} c_{\kp \sa}\ran \llan f_\bsa^\dag c_{\kpp\bsa} c_{ k \sa}\rran-\lan f_{\sa}^{\dag} c_{k \sa}\ran \llan f_\bsa^\dag c_{\kpp\bsa} c_{ \kp \sa}\rran,\\
\llan n_\bsa c_{\kpp\sa}^{\dag} c_{\kp \sa}c_{ k \sa}\rran&\approx& f_{\kpp\kp}^{\sa} \llan n_\bsa c_{ k \sa}\rran-f_{\kpp k}^{\sa} \llan n_\bsa c_{ \kp \sa}\rran,\\
\llan c_{\kpp\bsa}^\dag f_\bsa c_{\kp \sa}^{\dag}c_{ k \sa}f_{\sa}\rran&\approx& f_{\kp k}^{\sa}\llan c_{\kpp\bsa}^\dag f_\bsa f_{\sa}\rran-\lan c_{\kp \sa}^{\dag} f_{\sa}\ran \llan c_{\kpp\bsa}^\dag f_\bsa c_{ k \sa}\rran,\\
\llan f_\bsa^\dag c_{\kpp\bsa}c_{\kp \sa}^{\dag}c_{ k \sa}f_{\sa}\rran&\approx& f_{\kp k}^{\sa}\llan f_\bsa^\dag c_{\kpp\bsa} f_{\sa}\rran-\lan c_{\kp \sa}^{\dag} f_{\sa}\ran \llan f_\bsa^\dag c_{\kpp\bsa} c_{ k \sa}\rran.
\end{subeqnarray}
Again the Green functions $\llan  n_\bsa n_\sa c_{ k\sa}\rran$ vanish in the wide-band limit and are removed hereafter. Using Eqs.~(\ref{eq:f4}, \ref{eq:f4a}), Eqs.~(\ref{eq:nc03}) become
 \begin{subeqnarray}
\label{eq:nc04}
(\wa_{\sa:k \kp}+U)\llan n_\bsa f_{\sa}^{\dag} c_{\kp \sa}c_{ k \sa}\rran &=& \Big[\sum_\kpp  t_\sa   f_{\kpp k}^\sa -\wa_{:\kp} \lan f_{\sa}^{\dag} c_{k \sa}\ran\Big]\llan n_\bsa c_{\kp \sa}\rran -\Big[ \sum_\kpp t_\sa f_{\kpp\kp}^\sa-\wa_{:k} \lan f_{\sa}^{\dag} c_{\kp \sa}\ran\Big] \llan n_\bsa c_{ k \sa}\rran ,\slabel{eq:nc041}\\
(\wa_{\kp:\sa k}-U)\llan  n_\bsa c_{\kp \sa}^{\dag}c_{ k \sa}f_{\sa}\rran &=&f_{\kp k}^\sa(\wa_{:\sa} -U)\langle\langle n_\bsa f_{\sa}\rran -\Big[\sum_\kpp t_\sa f_{\kp \kpp}^\sa+\wa_{:k}\lan  c_{\kp \sa}^{\dag}f_{\sa}\ran\Big] \llan  n_\bsa c_{ k \sa} \rran.\slabel{eq:nc042}
\end{subeqnarray}
Combining Eqs.~(\ref{eq:nc011},~\ref{eq:nc04}) and using Eq.~(\ref{eq:acc03}) yield Eq.~(\ref{eq:f4d03}).

\renewcommand{\theequation}{\mbox{B.\arabic{equation}}} 
\setcounter{equation}{0} 

\section{DERIVATION OF EXPECTATION VALUES\label{App:TA1}}

At equilibrium, the hermiticity of expectation values holds, e.g. $\lan c^\dag_{\kp\sa} c_{k\sa}\ran=\lan c_{k\sa}^\dag c_{\kp\sa} \ran, ~\langle f_\sa^{\dag} c_{k\sa}\rangle=\langle c_{k\sa}^{\dag}f_\sa \rangle $ because $\mathcal{G}_\sa^<(\wa)=-f_F(\wa) [\mathcal{G}_\sa^r(\wa)-\mathcal{G}_\sa^a(\wa)]$. In the Keldysh formalism\cite{Langreth1972}, $\mathcal{G}_\sa^<(\wa)$ is the {\it lesser} Green function, while $\mathcal{G}_\sa^{r(a)}(\wa)$ is the retarded (advanced) Green function. This relation is nothing but the spectral theorem or equivalently Eq.~(\ref{eq:SF01}), with which it is standard to transform expectation values into a functional of the retarded or advanced dot Green function\cite{Kashcheyevs2006}. However, the spectral theorem does not apply out of equilibrium\cite{MW92} and it is therefore necessary to invoke the nonequilibrium Keldysh formalism\cite{Langreth1972}. We show how to rewrite the expectation values in Eq.~(\ref{eq:GF_01}) in terms of integral functions. We calculate for our purposes the following expectation values:
\bea
\lan f^\dag_\sa c_\ka \ran &\equiv&-i \int \frac{d\wa}{2\pi}\mathcal{G}^<_{\ka,\sa}(\wa)=-i t_\sa\int \frac{d\wa}{2\pi}\left[g_\ka^r(\wa)\mathcal{G}^<_{\sa}(\wa)+g_\ka^<(\wa)\mathcal{G}^a_{\sa}(\wa)\right]\nn\\
&=&t_{\sa}\left[ f_F(\va_k) \mathcal{G}_{\sa}^a(\va_k) + \int \frac{d\wa}{2\pi i}\frac{\mathcal{G}_{\sa}^<(\wa)}{\wa_{:k}+i \delta}\right] ,  \label{eq:e12}\\
\lan  c^\dag_\ka f_\sa\ran &\equiv&-i \int \frac{d\wa}{2\pi}\mathcal{G}^<_{\sa,\ka}(\wa)= -i t_\sa\int \frac{d\wa}{2\pi}\left[\mathcal{G}^r_{\sa}(\wa)g_\ka^<(\wa)+\mathcal{G}^<_{\sa}(\wa)g_\ka^a(\wa)\right]\nn\\
&=&t_\sa\left[ f_F(\va_k) \mathcal{G}_{\sa}^r(\va_k) + \int \frac{d\wa}{2\pi i}\frac{\mathcal{G}_{\sa}^<(\wa)}{\wa_{:k}-i \delta}\right],\\
\lan c^\dag_\kb c_\ka \ran &\equiv&-i \int \frac{d\wa}{2\pi}\mathcal{G}^{< }_{\ka,\kb}(\wa) \nn\\
&=& -i \int \frac{d\wa}{2\pi} \Big\{\delta_{k\kp}    g_\ka^{< }(\wa)  +t^2_\sa  \left[g^r_\ka(\wa)\mathcal{G}^r_{\sa}(\wa)g_\kb^{< }(\wa)\right.\nn\\
&&\left.+g^r_\ka(\wa)\mathcal{G}^{< }_{\sa}(\wa)g_\kb^a(\wa)+ g^{< }_\ka(\wa)\mathcal{G}^a_{\sa}(\wa)g_\kb^a(\wa) \right]\Big\}\nn\\
&=& \delta_{k\kp} f_F(\va_k) + t^2_\sa\Big[\frac{ f_F(\va_k) \mathcal{G}^a_{\sa}(\va_k) - f_F(\va_\kp)\mathcal{G}^r_{\sa}(\va_\kp)}{\va_k-\va_\kp -i \delta}
+ \int \frac{d\wa}{2\pi i}\frac{\mathcal{G}_{\sa}^<(\wa)}{(\wa_{:k}+i \delta)(\wa_{:\kp}-i \delta)} \Big]. \label{eq:e14}
\eea

where $g^{r(a)}_\ka(\wa)=(\wa_{:k}\pm i\delta)^{-1}$, $g_\ka^{<}(\wa)=2\pi i f_F(\va_k)\delta(\wa_{:k})$ are the bare lead Green functions. Using Eqs.~(\ref{eq:e12}-\ref{eq:e14}), we obtain after summation two related functions defined by $P_\sa(\wa)$ and $Q_\sa(\wa)$
\begin{align}
P_\sa(\wa)&\equiv\sum_{k}\frac{t_\sa \langle f_\sa^{\dag} c_{k\sa}\rangle}{\wa-\va_k+i\delta}=\sum_{\alpha=L,R}\frac{\Gamma_{\alpha\sa}}{\pi }\int d\va \frac{f^\alpha_F(\va)\mathcal{G}_\sa^{a}(\va)}{\wa-\va+i\delta},  \label{eq:taP}\\
Q_\sa(\wa)&\equiv\sum_{k\kp }\frac{t_\sa^2\lan c^\dag_{\kp\sa}
c_{k\sa}\ran}{\wa-\va_k+i\delta}=\sum_{\alpha=L,R}\frac{\Gamma_{\alpha\sa}}{\pi
}\int d\va
\frac{f^\alpha_F(\va)[1+i\Gamma_\sa\mathcal{G}_\sa^{a}(\va)]}{\wa-\va+i\delta}
 . \label{eq:taQ}
\end{align}

Notice that the imaginary part of the denominator is always positive, so that the pole $\va=\wa+i\delta$ remains in the upper half complex plane. In deriving Eqs.~(\ref{eq:taP}-\ref{eq:taQ}), some terms, particularly those associated with $\mathcal{G}^{< }_{\sa}(\wa)$, vanish in the wide-band limit, since upon summing over $k$, all denominators have poles in the upper half complex plane. Hence we have shown that in the wide-band limit, the nonequilibrium functions $P_\sa(\wa), ~Q_\sa(\wa)$ take the same forms as in equilibrium, except that the left and right leads have different chemical potentials. No knowledge of lesser Green functions is needed. This constitutes a huge simplification in the computations.

\renewcommand{\theequation}{\mbox{C.\arabic{equation}}} 
\setcounter{equation}{0} 

\section{Charge conjugation symmetry\label{App:CCS}}

This appendix is devoted to proving charge conjugation symmetry of the dot Green function given by Eq. (\ref{eq:GF_01}). We follow the scheme established by V. Kashcheyevs {\it et al}\cite{Kashcheyevs2006} who proved that this identity holds for the Lacroix approximation. The Anderson Hamiltonian (\ref{AndersonModel}) attains its original structure if replacing the particle operators by the hole ones, $\tilde{f}_\sa^\dag\equiv f_\sa,~\tilde{c}_{k\sa}^\dag\equiv c_{k\sa}$, along with
\bd
\mathcal{C}(\va_\sa) =-\va_\sa-U,\quad\mathcal{C}(U)=U,\quad
\mathcal{C}(t_\sa)=-t_\sa^*,\quad \mathcal{C}(\va_k)=-\va_k,\quad \mathcal{C}(\lan n_\sa \ran) =1- \lan n_\sa \ran, \label{PHtrans}
\ed
where $\mathcal{C}$ is the charge conjugation operator transforming electrons quantities into hole ones and reversely. The hole dot Green function is related to the particle dot Green function by charge conjugation symmetry
\bea
\mathcal{C}\left[\mathcal{G}^{r}_\sa(\wa)\right] \equiv \llan \mathcal{C}(f_\sa), \mathcal{C}( f_\sa^\dag)\rran_\wa=-\mathcal{G}_\sa(-\wa).\label{CCS_transform01}
\eea
Eq.~(\ref{eq:GF_01}) obeys this symmetry if the following rules are respected:
\begin{align}
\mathcal{C}\left[u_{1\sa}(\wa)\right]&=U-u_{1\sa}(-\wa), & \mathcal{C}\left[u_{2\sa}(\wa)\right]&=1-u_{2\sa}(-\wa),\nn\\
\mathcal{C}\left[\Sigma_\sa^0(\wa)\right]&=-\Sigma_\sa^0(-\wa),&
\mathcal{C}\left[P_\sa(\wa)\right]&=-P_\sa(-\wa),\label{CCS_transform}\\
\mathcal{C}\left[Q_\sa(\wa)\right]&=Q_\sa(-\wa)-\Sigma^0_\sa(-\wa).\nn
\end{align}
Therefore we deduce that $\mathcal{C}\left[P_\sa(\wa_{\bsa:\sa})\right]=-P_\sa(-\wa_{\sa:\bsa}),~\mathcal{C}\left[Q_\sa(\wa_{\bsa:\sa})\right]=Q_\sa(-\wa_{\sa:\bsa})-\Sigma_\sa^0(-\wa_{\sa:\bsa})$, and similarly $\mathcal{C}\left[P_\sa(-\wa_{:\sa\bsa}+U)\right]=-P_\sa(\wa_{\sa\bsa:}+U),~\mathcal{C}\left[Q_\sa(-\wa_{:\sa\bsa}+U)\right]=Q_\sa(\wa_{\sa\bsa:}+U)-\Sigma_\sa^0(\wa_{\sa\bsa:}+U)$. For the Lacroix approximation, this is enough to prove that the Green function respects charge conjugation symmetry $\mathcal{C}\left[\mathcal{G}^{r}_\sa(\wa)\right]=-\mathcal{G}_\sa(-\wa)$\cite{Kashcheyevs2006}. However, Eq.~(\ref{eq:GF_01}) has additional self-energy corrections in the arguments of $P_\sa(\wa)$ and $Q_\sa(\wa)$. Here we will show Eq.~(\ref{CCS_transform}) still holds for Eq.~(\ref{eq:GF_01}).
Using Eq.~(\ref{PHtrans}), we find
\begin{align}
\mathcal{C}\left[\wa_{\sa:k\kp}\right]&=\wa_{k\kp:\sa} -U,& \mathcal{C}\left[\wa_{\kp:\sa k}\right]&=\wa_{\sa k:\kp} + U,\nn \\
\mathcal{C}\left[D_{\sa:k\kp}\right]&=  D_{\sa:k\kp}|_{\wa\rightarrow-\wa},&  \mathcal{C}\left[D_{k:\sa\kp}\right]&=  D_{k:\sa\kp}|_{\wa\rightarrow-\wa},\nn\\
\mathcal{C}\left[\widehat{\Sigma}_2(\wa_{:k})\right]&=-\widehat{\Sigma}_{2\sa}(-\wa_{k:}),&   \mathcal{C}\left[\widehat{\Sigma}_3(\wa_{k:})\right] &=-\widehat{\Sigma}_{3\sa}(-\wa_{:k}).\nn
\end{align}
Thus if we redefine
\begin{align}
Q_\bsa(\wa_{\bsa:\sa})&\equiv\sum_{k\kp}\frac{t_\bsa^2f_{\kp k}^\bsa}{\wa_{\bsa:k\sa}-\widehat{\Sigma}_{2\sa}(\wa_{:k})},&
Q_\bsa(-w_{:\sa\bsa}+U)&\equiv\sum_{k\kp}\frac{t_\bsa^2 f_{k \kp}^\bsa}{-w_{k:\sa\bsa}+U+\widehat{\Sigma}_{3\sa}(\wa_{k:})},\\
P_\bsa(\wa_{\bsa:\sa})&\equiv\sum_k\frac{t_\bsa\langle f_\bsa^{\dag} c_{k\bsa}\rangle}{\wa_{\bsa:k\sa}-\widehat{\Sigma}_{2\sa}(\wa_{:k})},&
P_\bsa(-w_{:\sa\bsa}+U)&\equiv\sum_k\frac{t_\bsa\langle c_{k\bsa}^{\dag} f_\bsa\rangle}{-w_{k:\sa\bsa}+U+\widehat{\Sigma}_{3\sa}(\wa_{k:})}.
\end{align}
One can see that the above functions obey the above transformation rules (\ref{CCS_transform}) by analysing the transformation rules of their self-energies $\Sigma_{i\sa}$. In Eq.~(\ref{eq:GF_01}), there are yet another two terms, defined by
\begin{align}
Q_{1\sa}(\wa)&=\sum_{k\kp}  \frac{t_{\sa}^2f_{\kp k}^\sa\Sigma_{5\bsa}(\wa_{:k})}{[\wa_{:k}-\Sigma_{6\sa}(\wa_{:k})][\wa_{:k}-\Sigma_{1\bsa}(\wa_{:k})]},& 
P_{1\sa}(\wa)&=\sum_{k\kp}  \frac{t_{\sa}\Sigma_{5\bsa}(\wa_{:k})\lan f^\dag_\sa c_{k\sa}\ran}{[\wa_{:k}-\Sigma_{6\sa}(\wa_{:k})][\wa_{:k}-\Sigma_{1\bsa}(\wa_{:k})]}.
\end{align}
One can show that
\begin{align}
\mathcal{C}\left[P_{1\sa}(\wa)\right]&=-P_{1\sa}(-\wa),\nn \\
\mathcal{C}\left[Q_{1\sa}(\wa)\right]&=Q_{1\sa}(-\wa)-\Sigma^0_\sa(-\wa)\nn
\end{align}
in the wide-band limit. With this assumption, performing Eq.~(\ref{PHtrans}) on Eqs.~(\ref{eq:SE001},\ref{eq:SE005}) yields
\[
\mathcal{C}\left[\Sigma_{1\sa}(\wa_{:k})\right]=-\Sigma_{1\sa}(-\wa_{k:}) ,\quad \mathcal{C}\left[\Sigma_{5\sa}(\wa_{:k})\right]=-\Sigma_{5\sa}(-\wa_{k:}) ,\quad \mathcal{C}\left[\Sigma_{6}(\wa_{:k})\right]=-\Sigma_{6}(-\wa_{k:}).
\]
Therefore Eq.~(\ref{CCS_transform}) holds for $Q_{1\sa}(\wa)$ and $P_{1\sa}(\wa)$; similarly for $u_{1\sa}(\wa)$ and $u_{2\sa}(\wa)$. As a result, we prove that Eq.~(\ref{eq:GF_01}) maintains charge conjugation symmetry by obeying the particle-hole relation Eq.~(\ref{CCS_transform01}).

\end{widetext}

\renewcommand{\theequation}{\mbox{E.\arabic{equation}}} 
\setcounter{equation}{0}

\section{Symmetry with respect to the particle-hole symmetric point\label{App:symmH}}

In the previous section, we showed that the Green functions calculated from electron and hole Hamiltonians are related by charge conjugation symmetry. By changing $f_\sa\rightarrow \mathcal{C}(f_\sa^\dag),~c_{k\sa}\rightarrow \mathcal{C}(c_{k\sa}^\dag)$, the Anderson Hamiltonian for holes $\mathcal{C}\left[\mathcal{H}\right] = \mathcal{H}_h$ is, within a constant energy,
\begin{widetext}
\bea
  \mathcal{H}_h &=&   \sum_{\alpha \sa k} (- \va_{\alpha k}) \mathcal{C}(c_{\alpha k\sa}^{\dag}) \mathcal{C}(c_{\alpha k\sa}) + \sum_{\sa} \left(-\va_{\sa}-U\right) \mathcal{C}(n_{\sa}) + U \mathcal{C}(n_{\uparrow}) \mathcal{C}(n_{\downarrow}) - \sum_{\alpha \sa k} \left(t_{ \alpha \sa} \mathcal{C}(c^{\dag}_{\alpha k\sa}) \mathcal{C}(f_{\sa}) + H.C.\right), \slabel{hole_H}
\eea
\end{widetext}
where we keep the parameters of the original Hamiltonian for electrons. It maintains the structure of an Anderson Hamiltonian, but with transformed hole operators. \

\begin{figure}
    \centering
        \includegraphics[width=0.50\textwidth]{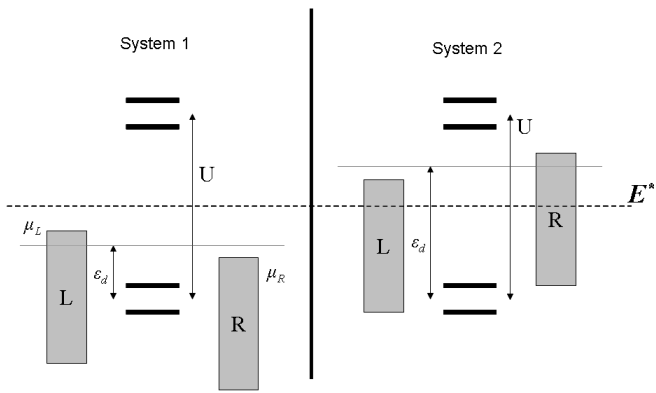}
    \caption{Schematic representation of two dual systems in the electron picture. The dotted line represents the particle-hole symmetric point $E^*$. Each system is equal to the mirror symmetry of its dual counterpart around $E^*$ in energy space . The electrons in System 1 must behave exactly the same as the holes in System 2, and reversely. \
}
    \label{fig:cartoon_p-h_symm_point}
\end{figure}

Here lies another symmetry: to the hole Hamiltonian $\mathcal{H}_h$ (\ref{hole_H}) corresponds an electron Hamiltonian $\mathcal{H}_e^{\text{Syst2}}$ (\ref{hole_e}) of another system, whose parameters share with $\mathcal{H}_h$:
\begin{widetext}
\bea
  \mathcal{H}_e^{\text{Syst2}}&=&   \sum_{\alpha \sa k} ( \va_{ k}+\mu_\alpha) c_{\alpha k\sa}^{\dag} c_{\alpha k\sa} + \sum_{\sa} \left(-\va_{\sa}-U\right)n_{\sa} + U n_{\uparrow} n_{\downarrow} - \sum_{\alpha \sa k} \left(t_{ \alpha \sa} c^{\dag}_{\alpha k\sa}f_{\sa} + H.C.\right).
  \slabel{hole_e}
\eea
\end{widetext}
 We call dual systems two systems showing the symmetry $\mathcal{H}_h^{\text{Syst1}} = \mathcal{H}_e^{\text{Syst2}}$, as shown in Fig. \ref{fig:cartoon_p-h_symm_point}. For instance, one can have the following parameters\
\begin{center}
        \begin{tabular}{c|c|c}
            & System 1 & System 2 \\
            \hline
            $\mathcal{H}_e$ & $\quad \va_d=-2$ ; $U=6 \quad$ & $\quad \va_d=-4$ ; $U=6$ \\
            $\mathcal{H}_h$ & $\quad \mathcal{C}(\va_d)=-4$ ; $\mathcal{C}(U)=6 \quad$ & $\quad \mathcal{C}(\va_d)=-2$ ; $\mathcal{C}(U)=6$     \end{tabular}
            \end{center}
            An electron in the first system behaves exactly as a hole in the second (dual) system and reversely.\

This symmetry is slightly broken by our approximation scheme; the
worst case is in the $\mathcal{N}=1$ CB regime. The reason could
be due to the fact that at order $t_\sa^4$ we do not treat the
particle and hole contributions on an equal footing. Therefore,
because the transition rates of the self-energies
$\widehat{\Sigma}_{2\sa}$ and $\widehat{\Sigma}_{3\sa}$ have
different values in the frequency range $\va_\sa\le \wa\le
\va_\sa+U$, it leads to slightly asymmetric renormalization and
broadening of the resonant peaks at $\va_\sa$ and $\va_\sa+U$, as
can be shown in the particle-hole symmetric case, which affects
the occupation number. For instance, at the particle-hole
symmetric point ($\va_\sa=-U/2$, $\mu_L=-\mu_R$), the dot
occupation number $\lan n_\sa \ran $ is expected  to be exactly
$1/2$ in equilibrium or in the symmetric bias setting. Our
numerical result shows deviation by a few percents at worst.
However, it has almost no effect on the low-frequency density of states
structure.\

In order to restore the symmetry, one compute the Green function in the dual system. Because of the definition of the duality, we have the identity
\bea
\mathcal{G}^{\text{Syst1}}_\sa(\wa) = \mathcal{C}\left[(\mathcal{G}^{\text{Syst2}}_\sa(\wa)\right] ,
\eea
where Systems 1 and 2 are dual of each other. Using charge conjugation symmetry (cf.\ref{App:CCS}) on System 2, we can express this equality in terms of electron Green functions only, that is
\bea
\mathcal{G}^{\text{Syst1}}_\sa(\wa) = - \left[\mathcal{G}^{\text{Syst2}}_\sa(-\wa)\right]^* . \nn
\eea
As mentioned earlier, this equality is slightly violated at high frequencies by our approximation scheme. We therefore symmetrise the two by setting
\bea
\mathcal{G}^{r}_\sa(\wa) = \left\{\mathcal{G}^{\text{Syst1}}_\sa(\wa)  - \left[\mathcal{G}^{\text{Syst2}}_\sa(-\wa)\right]^*\right\}/2 .
\eea

\end{document}